\RequirePackage{fix-cm}
\documentclass[smallextended]{svjour3}       
\smartqed  
\usepackage{graphicx}
\usepackage{algorithm}
\usepackage{amsmath}
\usepackage{amssymb}
\usepackage{xcolor}
\usepackage{latexsym}
\usepackage{float}
\usepackage{epsfig,subfigure,epstopdf,graphicx,array}
\usepackage{algpseudocode}
\usepackage[]{graphicx}

\begin{document}

\title{Low-Complexity Interference Cancellation Algorithms for Detection in Media-based Modulated Uplink Massive-MIMO Systems
}

\author{Manish~Mandloi* \and Devendra~Singh~Gurjar}


\institute{M. Mandloi (Corresponding author)  \at
              Department of Electronics and Telecommunication Engineering, SVKM's NMIMS (Deemed to be University) Shirpur campus, Maharashtra 425405, India\\
              Tel.: +91-9584858734\\
              \email{manish.mandloi@nmims.edu}           
           \and
           D. S. Gurjar \at
           Department of Electronics and Communication Engineering, National Institute of Technology Silchar, Assam 788010, India\\
           \email{dsgurjar@ece.nits.ac.in}
}

\date{Received: date / Accepted: date}

\maketitle

\begin{abstract}
Media-based modulation (MBM) is a novel modulation technique that can improve the spectral efficiency of the existing wireless systems. In MBM, multiple radio frequency (RF) mirrors are placed near the transmit antenna(s) and are switched ON/OFF to create different channel fade realizations. In such systems, additional information is conveyed through the ON/OFF status of RF mirrors along with conventional modulation symbols. A challenging task at the receiver is to detect the transmitted information symbols and extract the additional information from the channel fade realization used for transmission. In this paper, we consider a massive MIMO (mMIMO) system where each user relies on MBM for transmitting information to the base station, and investigate the problem of symbol detection at the base station. First, we propose a mirror activation pattern (MAP) selection based modified iterative sequential detection algorithm. With the proposed algorithm, the most favorable MAP is selected, followed by the detection of symbol corresponding to the selected MAP. Each solution is subjected to the reliability check before getting the update. Next, we introduce a $K$ favorable MAP search based iterative interference cancellation (KMAP-IIC) algorithm. In particular, a selection rule is introduced in KMAP-IIC for deciding the set of favorable MAPs over which iterative interference cancellation is performed, followed by a greedy update scheme for detecting the MBM symbols corresponding to each user. Simulation results show that the proposed detection algorithms exhibit superior performance-complexity trade-off over the existing detection techniques in MBM-mMIMO systems.
\end{abstract}

\section{Introduction}
\label{sec1}
The explosive increase in the number of subscribers, the use of data thirsty applications, and ubiquitous computing for low latency applications pose a serious challenge towards the design of advanced communication techniques for 5G and beyond wireless systems. Over the last decade, different wireless techniques, such as non-orthogonal multiple access (NOMA), millimeter-wave (mm-Wave) communications, vehicle-to-everything (V2X) communications, massive MIMO (mMIMO), and machine learning in wireless networks, have been proposed in the literature to enhance the performance of the existing wireless systems \cite{r1b,r1c,r1a}. Amongst these, mMIMO has been considered as a promising technique to satisfy the requirement of high data rate for 5G and beyond wireless systems \cite{r1,r2,r3}. In mMIMO, a large number of base station (BS) antennas are used to serve comparatively small number of single/multiple antenna users \cite{r1,r2}. One of the key advantages of such systems is that simple linear precoders and decoders can achieve near-optimal bit error rate (BER) performance which includes zero-forcing and minimum mean squared error detectors \cite{r1,r4}. Some of the key challenges in practical implementation of multiple antenna systems include inter channel interference, requirement of dedicated RF chains at each transmit antenna, and constraints on the total number of receive antennas. Index modulation is one such digital modulation technique proposed to overcome these challenging issues in MIMO systems \cite{r5,r5a}. Through index modulation, extra information bits are embedded in the switching pattern of the building blocks along with the transmission of symbols selected from the conventional constellation set such as M-ary PSK or M-QAM \cite{r6}. This results in relaxing the need of all the available resources at the transmitter such as transmit antennas and RF chains \cite{r7,r8}. In conventional index modulation (IM) schemes, such as spatial modulation (SM) and generalised SM (GSM), only one or fewer antenna elements and RF chains are active which reduces the energy consumption as well as the inter channel interference thereby providing a higher energy efficiency. IM together with mMIMO \cite{r9,r10} has been evolved as an emerging technique for fulfilling the higher spectral as well as the higher energy efficiency requirements in beyond fifth-generation (B5G) wireless systems \cite{r6}. One of the key challenges in IM-mMIMO systems is the detection of the transmitted information which requires detection of the selected switching pattern and the detection of transmitted symbol.\\
\indent Recently, media-based modulation (MBM) has been introduced as a potential IM scheme to enhance the spectral efficiency and the energy efficiency of the existing wireless systems by embedding information in different channel fade realizations \cite{r11}, \cite{r12}. These different end-to-end channel fade realizations are created by modifying the radio frequency (RF) properties of the propagation medium near to the transmitter \cite{r11}. To achieve these variations, multiple RF mirrors (parasitic elements), which are digitally controlled by the input data bits, are placed near the transmit antenna(s) \cite{r12}--\cite{Choc} . The ON/OFF status of RF mirrors also referred to as mirror activation pattern (MAP) convey additional information bits along with the conventional modulation symbol. Through $n_{rf}$ RF mirrors, $2^{n_{rf}}$ MAPs can be generated. Each MAP results in an independent channel fade realization which is used by the transmitting antenna to transmit information symbol selected from a constellation set. The practical advantages of MBM systems have been demonstrated in \cite{r12} with 14 RF mirrors placed near a dipole transmit antenna. Moreover, the antenna design, radiation pattern, and potential advantages of MBM are discussed in \cite{r12}, \cite{r1111}. The use of MBM in mMIMO systems, which is referred as MBM-mMIMO, has been manifested to achieve significant improvements in terms of BER performance, and spectral efficiency \cite{r13,r14} over the existing modulation schemes such as SM and GSM \cite{r15,r16,r17,r18,r19,r20}. It is worth noting that, in MBM the throughput enhancement is linear with respect to the number of RF mirrors, whereas, in SM and GSM, the increase in throughput is logarithmic \cite{r13}. Therefore, MBM is being considered as an emerging spectral efficient IM scheme in B5G wireless systems. However, one of the major challenges in MBM-mMIMO systems is to detect the transmitted information symbol, and to extract the additional information conveyed through the selection of channel fade realization by each user, reliably with low computational complexity \cite{r13,r21}.\\ 
\indent Optimal detection rule suggests an exhaustive search over the set of all the possible MBM symbol vector \cite{r13,r14,r21}, which is termed as maximum-likelihood (ML) detection. However, due to exponentially high computational complexity, ML detection is impractical in MBM-mMIMO systems \cite{r13}. It is due to the fact that the set of all the possible symbol vectors for each user is generated by using all the possible combinations of MAPs and constellation points which ultimately results in an exponential increase in the size of the search space for ML detection with respect to the number of transmit antennas and $n_{rf}$. Though, ZF/MMSE achieves near-optimal performance in mMIMO systems \cite{r1}, \cite{r22,r23,r24,r25}, their performance in MBM-mMIMO systems is sub-optimal which makes ZF/MMSE less selective for MBM-mMIMO systems \cite{r7,r8}. The sub-optimal BER performance of ZF/MMSE and exponentially high computational complexity of ML detection motivates for the design of low-complexity detection algorithms capable of achieving better BER performance in MBM-mMIMO systems.\\
\indent Recently, inclusion-exclusion subspace pursuit (IESP) \cite{r13} and iterative interference cancellation (IIC) \cite{r14} algorithms have been proposed to achieve better BER performance over the conventional detection techniques such as MMSE and successive interference cancellation (SIC) \cite{r12,r14,r26} in MBM-mMIMO systems. However, the IESP algorithm \cite{r13} requires computation of pseudo inverse at multiple stages which is computationally expensive in MBM-mMIMO system due to large dimensional channel matrix. On the other hand, IIC algorithm \cite{r14} outperforms MMSE, SIC and IESP algorithms in terms of both the BER performance and the computational complexity. In IIC \cite{r14}, symbol vector transmitted by each user is detected by canceling interference from all the other users followed by an exhaustive search over the set of all the possible MAPs in an iterative manner. Although the computational complexity of IIC is less compared with MMSE, SIC and IESP algorithms, there is scope to reduce the computational complexity of IIC algorithm further by selecting a list of favorable MAPs for performing the search. In order to solve the detection problem in MBM-mMIMO systems, we propose two low-complexity detection schemes in this paper. First, we propose a MAP selection based iterative sequential detection (MAP-ISD) algorithm where the most favorable MAP is selected using a selection metric. The symbol corresponding to the selected MAP is detected sequentially for each user while nullifying the interference from all the other users. Next, we propose a $K$ favorable MAP search based iterative interference cancellation (KMAP-IIC) algorithm. In particular, $K$ most favorable MAPs are selected over which the search is performed for detecting the symbol corresponding to each user. The selection metric is obtained by exploiting the \textit{channel hardening} which occurs in mMIMO systems \cite{r27}.\\
Our key contributions in this article are;
\begin{itemize}
\item A technique for selection of favorable mirror activation pattern (MAP) for each user is proposed by utilizing channel hardening in mMIMO.
\item A low-complexity MAP-ISD algorithm is proposed for symbol detection in the MBM-mMIMO systems which utilizes the concept of favorable MAP and low-complexity ML search in sparse vectors. 
\item The concept of reliability check of the 
solution in each iteration and stopping rule are utilized in the MAP-ISD algorithm to obtain better BER with low computational complexity.
\item Selection of $K$ favorable MAPs and low-complexity ML search are also integrated with IIC and proposed KMAP-IIC algorithm.
\item Simulation results on BER and computational complexity of the proposed algorithms are shown to validate superiority of the proposed algorithms over the existing algorithms.
\end{itemize}

Simulation results show that the proposed detection algorithms exhibit superior performance-complexity trade-off over the existing detection techniques such as MMSE, IESP, and IIC \cite{r13,r14} in MBM-mMIMO systems. It is also observed through simulations that the selection of a detection scheme between MAP-ISD and KMAP-IIC depends on the number of RF mirrors and the number of users in the system.\\ 
\textit{Notation:} Boldface upper-case and lower-case letters denote matrices and column vectors, respectively. $(\cdot)^H$ and $(\cdot)^{-1}$ represent matrix Hermitian and matrix-inversion, respectively. $x_i$ is the $i^{th}$ element of  $\textbf{x}$. $\textbf{I}_{U}$ refers to $U \times U$ identity matrix, $a_{i,j}$ denotes the element in $i^{th}$ row and $j^{th}$ column of  $\textbf{A}$, and $\textbf{a}_j$ denotes $j^{th}$ column of $\textbf{A}$. $\mathcal{Q}[\cdot]$ denotes the quantization operation which maps the soft values to the nearest constellation point.

\section{Background}
\label{sec2}
In this section, we discuss the mathematical model of MBM-mMIMO system, and introduce the ML detection rule for optimal detection of symbols in such systems. We also shed light on the prohibitive computational complexity of ML detection in MBM-mMIMO systems. Finally, we provide the detailed description of ISD and IIC algorithms for detection in mMIMO systems and MBM-mMIMO systems, respectively.
\begin{table}[H]
\caption{List of Notations}
\label{tab0}   
\centering    
\begin{tabular}{ll}

\noalign{\smallskip}\hline\noalign{\smallskip}
$N_r$ & Number of BS antennas  \\
\hline\noalign{\smallskip}
$U$ & Number of users \\
\hline\noalign{\smallskip}
$n_{rf}$ & Number of RF mirrors \\
\hline\noalign{\smallskip}
$M=2^{n_{rf}}$ & Number of mirror activation pattern (MAP) \\
\hline\noalign{\smallskip}
$\mathbb{A}$ & Constellation set \\
\hline\noalign{\smallskip}
$\mathbb{S}_{MBM}$ & MBM signal set for a single user\\
\hline\noalign{\smallskip}
$\mathbb{S}_{MBM}^U$ & MBM signal set for $U$ user system\\
\hline\noalign{\smallskip}
$L$ & Number of iterations \\

\noalign{\smallskip}\hline
\end{tabular}
\end{table}
\subsection{System model}
\label{ssec21}
\begin{figure}[H]
	\centering 
	\includegraphics[width = 10cm]{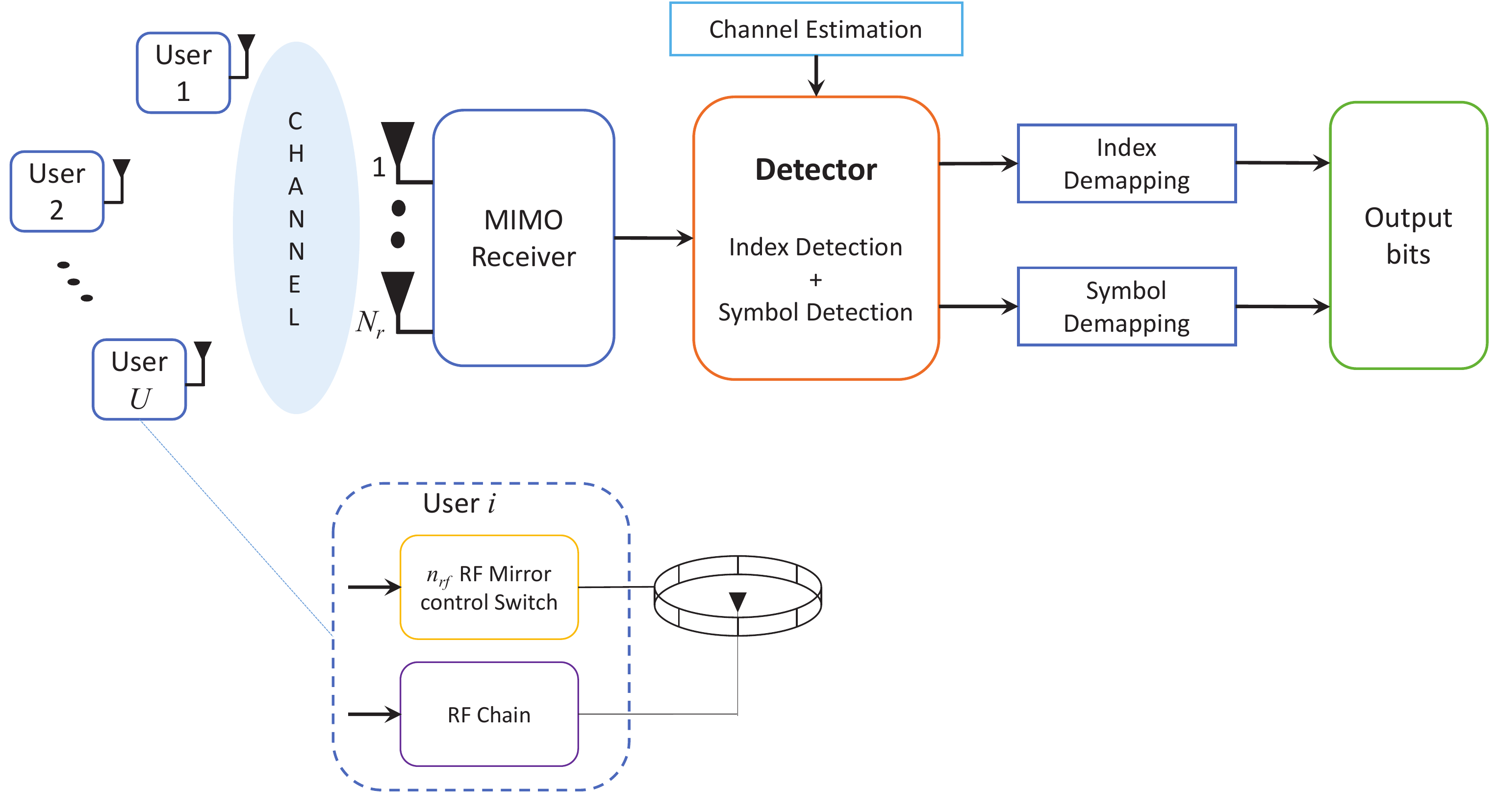}
	\caption{System model for MBM-mMIMO system.}
    \label{fig_sys}
\end{figure}
We consider an $N_r \times U$ mMIMO system with $N_r$ BS antennas and $U$ single antenna users ($N_r>>U$ for e.g., $N_r=128, U = 16$ ). Each user employs MBM for transmission of information to the BS as depicted in Fig. \ref{fig_sys}. We also consider that each user is having $n_{rf}$ RF mirrors placed near the antenna. Using MBM, the information is transmitted in two parts: 1. the mirror ON/OFF status, and 2. the conventional modulation symbol using the constellation set $\mathbb{A}$ (e.g., 4-QAM, QPSK, 16-QAM). To switch mirrors ON/OFF, each user requires $n_{rf}$ bits of incoming information, and therefore, there are $M = 2^{n_{rf}}$ ON/OFF combinations possible which are termed as MAPs. In a rich scattering environment, each MAP corresponds to an independent channel fade realization \cite{r11,r12}. After selection of an MAP, $\log_2{\vert{\mathbb{A}}\vert}$ additional bits are transmitted through the antenna by selecting one symbol from the modulation alphabet $\mathbb{A}$. Therefore, the spectral efficiency of a multi-user MBM system in terms of bits per channel use (bpcu) can be mathematically given by
\begin{equation}
\label{eq1}
\eta_{MBM} = U(n_{rf}+\log_2{\vert{\mathbb{A}}\vert}).
\end{equation}  
Let, the channel state vector between the $j$th MAP selected by $k$-th user and the base station is represented by $\mathbf{h}^j_k$. Each $\mathbf{h}^j_k=[h^j_{1,k}, h^j_{2,k}, \cdots, h^j_{N_r,k}]_{N_r \times 1}^T$, for all $j = 1,2,\cdots,M$ and $k = 1,2,\cdots,U$, where $h^j_{i,k}$ is assumed to be independent and identically distributed (i.i.d.) complex Gaussian random variable with zero mean and unit variance i.e. $\sim \mathcal{CN}(0,1)$. Therefore, the channel matrix comprising of channel vectors corresponding to all the possible MAPs for the $k$-th user is $\mathbf{H}_k =[\mathbf{h}^1_k, \mathbf{h}^2_k, \cdots, \mathbf{h}^M_k]$. The MBM signal set for a single user can be defined as the set of all the possible transmit symbol vectors as
\begin{gather}
\nonumber
\mathbb{S}_{MBM}=\{\mathbf{s}_{j,i}: j=1,2,\cdots,M, i=1,2,\cdots,\vert\mathbb{A}\vert\},\\
\label{eq2}
\text{s.t.}\ \ \mathbf{s}_{j,i}=[0,\cdots, 0, \underbrace{q_i}_{j\text{th coordinate}}, 0, \cdots, 0]^T, q_i\in\mathbb{A},
\end{gather} 
where $\mathbf{s}_{j,i}$ is an $M \times 1$ vector with only one non-zero entry $q_i \in \mathbb{A}$ corresponding to the $j$th channel fade realization. For the $k$-th user, let us denote the transmit symbol vector is denoted by $\mathbf{x}_k \in \mathbb{S}_{MBM}$. Thus, the received symbol vector at the BS after performing matched filtering and sampling operations can be written as
\begin{equation}
\label{eq3}
\mathbf{y}=\sum_{k=1}^{U} \mathbf{H}_k \mathbf{x}_k + \mathbf{n},
\end{equation}
where $\mathbf{n}$ is an $N_r \times 1$ additive white Gaussian noise (AWGN) vector with $\mathbf{n}\sim\mathcal{CN}(0,\sigma^2\mathbf{I}_{N_r})$. Further, for simplicity, we can rewrite the received symbol vector as
\begin{equation}
\label{eq4}
\mathbf{y}=\mathbf{H}\mathbf{x} + \mathbf{n},
\end{equation}
where $\mathbf{H}=[\mathbf{H}_1\ \  \mathbf{H}_2\ \  \cdots\ \  \mathbf{H}_U]$ is an $N_r \times UM$ MBM-mMIMO channel matrix and $\mathbf{x}=[\mathbf{x}_1^T\ \ \mathbf{x}_2^T\ \ \cdots\ \ \mathbf{x}_U^T]^T$ is $UM \times 1$ MBM-mMIMO transmit symbol vector comprising of transmit symbol vectors of all the users.

\subsection{Maximum Likelihood Detection for MBM-mMIMO systems}
\label{ssec22}
The objective at the receiver is to extract the information embedded in the selection of MAP as well as detect the transmitted information symbol from the constellation set for realizing the potential benefits of MBM. An optimal detection rule termed as ML detection suggests an exhaustive search over all the possible combinations of the transmit symbol vector $\mathbf{x}$ which minimize the ML cost, where ML cost associated with the symbol vector $\mathbf{x}$ is given by
\begin{eqnarray}
\label{eq5}
C(\mathbf{x}) & = & \Vert\mathbf{y}-\mathbf{H}\mathbf{x}\Vert^2,\\
\label{eq6}
& = & \Vert\mathbf{y}-\sum_{k=1}^{U} \mathbf{H}_k \mathbf{x}_k\Vert^2.
\end{eqnarray} 
Therefore, the ML detection problem given the received vector $\mathbf{y}$ and the channel state information $\mathbf{H}$ can be formulated as 
\begin{equation}
\label{eq7}
\hat{\mathbf{x}} = \arg \min_{\mathbf{x}\in \mathbb{S}^{U}_{MBM}} \Vert\mathbf{y}-\mathbf{H}\mathbf{x}\Vert^2.
\end{equation}
where $\mathbb{S}^{U}_{MBM}$ is the set of all the possible combinations of $\mathbf{x}$. In MBM-mMIMO systems, for a given value of $U$, $M$ and $\vert{\mathbb{A}}\vert$ the dimensions of $\mathbf{x}$ equals $M \times U$ and the set $\mathbb{S}^{U}_{MBM}$ contains $\left(M \times \vert{\mathbb{A}}\vert \right)^{U}$ possible combinations of $\mathbf{x}$ for e.g., if $U = 16$ and $M = 2^{n_{rf}} = 8$ then for a 4-QAM modulated MBM-mMIMO systems the dimensions of $\mathbf{x}$ is 128 ($M \times U = 16 \times 8 = 128$) and the size of $\vert \mathbb{S}^{U}_{MBM} \vert = 32^{16} \approx 1.2 \times 10^{24}$. Clearly, ML detection is unreasonable in such systems. Therefore, the design of low-complexity detection techniques is a challenging and a crucial problem for practical realization of MBM-mMIMO in the B5G wireless systems.

\subsection{Iterative Sequential Detection for mMIMO Systems}
\label{ssec23}
In this section, we discuss the ISD algorithm proposed in \cite{r24} for detecting the transmitted information symbols in mMIMO systems. In ISD, symbols corresponding to each user are detected in a sequential manner. In each iteration of ISD, symbols corresponding to all the users are detected which are refined in next iterations. An initial solution is used to initialize the algorithm, which could be an all zero solution, i.e., $\hat{\mathbf{x}}_j^{0}=0$ for all $j=1,2,\cdots,U$. Next, for detection of symbol corresponding to the $k$-th user in the $t$-th iteration, interference from all the other users is cancelled as
\begin{equation}
\label{eq8}
\mathbf{r}_k^{(t)} = \mathbf{y}-\sum_{i=1}^{k-1}\mathbf{h}_i\hat{\mathbf{x}}_i^{(t)}-\sum_{i=k+1}^{U}\mathbf{h}_i\hat{\mathbf{x}}_i^{(t-1)},
\end{equation}
where $\hat{\mathbf{x}}_i^{t-1}$ and $\hat{\mathbf{x}}_i^{t}$ is the symbol corresponding to the $i$-th user in the $(t-1)$-th and the $(t)$-th iterations, respectively. The residual error vector $\mathbf{r}_k^{(t)}$ is then passed through a matched filter for detecting $\hat{\mathbf{x}}_i^{t}$ as
\begin{equation}
\label{eq9}
\hat{\mathbf{x}}_i^{t} = \mathcal{Q}\left[\frac {\mathbf{h}_i^H}{\Vert \mathbf{h}_i \Vert ^2} \mathbf{r}_k^{(t)} \right],
\end{equation}
where $\mathcal{Q}[\cdot]$ is the quantization operation which maps the soft values to the nearest constellation points. These steps are repeated for multiple iterations to obtain a better solution. It is worth noting that the detection of MBM symbols using the ISD algorithm directly is an inefficient way of detection due to the presence of MAPs. Moreover, selecting a reliable solution for each user and a proper stopping rule are required to avoid error propagation and terminate the algorithm early, respectively. Therefore, the selection of a reliable MAP and solution for each user is crucial for detecting the transmitted symbol in MBM-mMIMO systems. Also, the stopping rule will help in terminating the algorithm early, thereby saving the required computations. This motivates further modifications in ISD to detect symbols in MBM-mMIMO systems with low-complexity. Algorithm \ref{algo1} presents the pseudo code of ISD algorithm for mMIMO detection.

\begin{algorithm}[H]
{
\caption{ISD Algorithm}
\label{algo1}
\small
\begin{algorithmic}[1]
\State {{\bf Input:} ${\mathbf y}$, ${\mathbf H}$, $N_r, U, L $}
\State {\bf Initialization:} $\mathbf{x}^{(0)}=\mathbf{0},\ \  \mathbf{r}^{(0)}=\mathbf{y},\ \  t = 0\ \  (\text{Iteration index})$\;
\Repeat
\State $\hat{\mathbf{x}}^{(t)}=\hat{\mathbf{x}}^{(t-1)}, \ \ \mathbf{r}^{(t)}=\mathbf{r}^{(t-1)}, \ \  t = t + 1$\;
\For {$(j=1 , ++j, j\leq U )$} 
\State Compute : $\mathbf{r}_k^{(t)} = \mathbf{y}-\sum_{i=1}^{k-1}\mathbf{h}_i\hat{\mathbf{x}}_i^{(t)}-\sum_{i=k+1}^{U}\mathbf{h}_i\hat{\mathbf{x}}_i^{(t-1)}$\;
\State Compute : $z_i = \frac {\mathbf{h}_i^H}{\Vert \mathbf{h}_i \Vert ^2}\mathbf{r}_k^{(t)}$\;
\State Update the symbol for $i$th user as : $\hat{\mathbf{x}}_i^{t} = \mathcal{Q}\left[ \mathbf{z}_i \right]$
\EndFor
\Until{$t=L$ {(terminate the iterative processing)}}
\State {\bf Output:}$\hat{{\mathbf{x}}}_{L}$
\end{algorithmic}
}
\end{algorithm}

\subsection{Iterative Interference Cancellation Algorithm for MBM-mMIMO Systems}
\label{ssec24}
In this section, we discuss the IIC algorithm proposed in \cite{r14}. In IIC, the symbol detection corresponding to each user is a two step process. The algorithm starts with an all zero solution, i.e., $\hat{\mathbf{x}}_i^{(0)}=0$ $\forall i=1,2,\cdots,U$. In the first step, the residual signal is obtained as

\begin{eqnarray}
\label{eq10}
\mathbf{r}^{(t)} & = & \mathbf{y}-\sum_{i=1}^{U}\mathbf{H}_i\hat{\mathbf{x}}_i^{(t)},\\
\label{eq11}
\mathbf{y}_j^{(t)} & = & \mathbf{r}^{(t)}+\mathbf{H}_j\hat{\mathbf{x}}_j^{(t)}
\end{eqnarray} 
where $\hat{\mathbf{x}}_i^{(t-1)}$ is the estimated MBM symbol vector of the $i$-th user in $(t-1)$-th iteration, and $\mathbf{r}^{(t)}$ is the residual signal. $\mathbf{y}_j^{(t)}$ is the interference cancellation vector for the $j$-th user in the $t$-th iteration. Next, a search is performed over the set of all the possible MBM symbol vectors for detecting the symbol vector corresponding to the $j$th user as
\begin{equation}
\label{eq12}
\hat{\mathbf{s}}_j^{(t)} = \arg\min_{\mathbf{s}_j\in \mathbb{S}_{MBM}} \Vert{\mathbf{y}_j^{(t)}-\mathbf{H}_j\mathbf{s}_j}\Vert^2.
\end{equation} 
IIC algorithm is initialised with an all-zero solution. After performing these two steps for all the users, a greedy search is performed multiple times to update the solution for each user by using the selection metric given as
\begin{equation}
\label{eq13}
u = \arg\min_{k=1,\cdots, U} \Vert{\mathbf{r}^{(t)}+\mathbf{H}_j(\hat{\mathbf{s}}_j^{(t)}-\hat{\mathbf{x}}_i^{(t)})}\Vert^2.
\end{equation}
These steps are performed for multiple iterations so that the algorithm converge to a better solution. The pseudo code of IIC algorithm is described in Algorithm \ref{algo2}.
\begin{algorithm}[H]
\caption{IIC Algorithm}
\label{algo2}
\small
\begin{algorithmic}[1]
\State {\bf Input:} ${\mathbf y}$, ${\mathbf H}$, $N_r, U, n_{rf}, L $
\State {\bf Initialization:} $\mathbf{x}^{(0)}=\mathbf{0}\ \  (\text{Symbol estimation}), \mathbf{r}^{(0)}=\mathbf{y}\ \  (\text{Residual signal}), \mathbf{W}_i, \forall i=1,2,\cdots,U, \ t = 0\ \  (\text{Iteration index})$\;
\Repeat
\State $\hat{\mathbf{x}}^{(t)}=\hat{\mathbf{x}}^{(t-1)}, \mathbf{r}^{(t)}=\mathbf{r}^{(t-1)}, \ \ t = t + 1$\;
\For {$(j=1 , ++j, j\leq U )$} 
\State Apply interference cancellation for the $j$-th user as: $\mathbf{y}_j^{(t)}  =  \mathbf{r}^{(t)}+\mathbf{H}_j\hat{\mathbf{x}}_j^{(t)}$ \;
\State Perform ML search for the $j$-th user as: $\hat{\mathbf{s}}_j^{(t)} = \arg\min_{\mathbf{s}_j\in \mathbb{S}_{MBM}} \Vert{\mathbf{y}_j^{(t)}-\mathbf{H}_j\mathbf{s}_j}\Vert^2$\;
\EndFor
\For {$(j=1 , ++j, j\leq U )$}
\State Select the first element of $u = \arg\min_{k=1,\cdots, U} \Vert{\mathbf{r}^{(t)}+\mathbf{H}_j(\hat{\mathbf{s}}_j^{(t)}-\hat{\mathbf{x}}_i^{(t)})}\Vert^2$\;
\If {$\Vert\mathbf{r}^{(t)}\Vert^2 > \Vert\mathbf{r}^{(t)}+\mathbf{H}_u(\hat{\mathbf{x}}_u^{(t)}-\hat{\mathbf{s}}_u^{(t)})\Vert^2$}
\State Update the residual error vector as $\mathbf{r}^{(t)} = \mathbf{r}^{(t)}+\mathbf{H}_u(\hat{\mathbf{x}}_u^{(t)}-\hat{\mathbf{s}}_u^{(t)})$\;
\State Update the symbol vector for $v$th user as $\hat{\mathbf{x}}_u^{(t)}=\hat{\mathbf{s}}_u^{(t)}$\;
\Else
\State break\;
\EndIf 
\EndFor
\Until{$\hat{\mathbf{x}}^{(t)}=\hat{\mathbf{x}}^{(t-1)}$ or $t=L$ {(terminate the iterative processing or if no update in the solution)}}
\State $\hat{\mathbf{x}}=\hat{\mathbf{x}}^{(t)}$
\State {\bf Output:}$\hat{{\mathbf{x}}}$
\end{algorithmic}
\end{algorithm}
The size of search space $\mathbb{S}_{MBM}$ for each user in MBM-mMIMO is $\left(M \times \vert \mathbb{A} \vert\right)$ which grows exponentially with the number of RF mirrors and linearly with the size of constellation set used. To perform ML search for each user in a single iteration, it requires IIC to search over $M \times \vert \mathbb{A} \vert$ possible solutions. Therefore, the total search operations in performing ML search is $\left(M \times \vert \mathbb{A} \vert\right)UL$. This makes the algorithm computationally expensive and motivates for further research in the algorithm to improve the practical feasibility of IIC in MBM-mMIMO systems.

\section{Proposed Detection Algorithms}
\label{sec3}
In this section, we discuss the proposed algorithms, namely, MAP-ISD algorithm and KMAP-IIC algorithm for MBM symbol detection in mMIMO systems.  

\subsection{MAP Selection based ISD}
In this algorithm, we propose to modify the ISD algorithm \cite{r24}, which was originally proposed to detect the symbols in mMIMO systems. Here, we improvise the ISD algorithm for detecting MBM symbols corresponding to each user in MBM-mMIMO systems. In particular, we introduce a rule for selecting the most favorable MAP corresponding to each user from the list of all the possible MAPs in MBM-mMIMO system. For this, first, we utilize the concept of \textit{channel hardening}, which occurs in mMIMO systems, and then find the pseudo-inverse of the individual channel matrices $\mathbf{H}_k$. The key idea for computing the pseudo-inverse is to find the highly erroneous locations by applying low-complexity zero-forcing over the residual vector (as discussed later in this section). The MAP corresponding to these erroneous locations are then explored for detecting the symbol transmitted by a particular user. Upon detection, the residual vector is updated accordingly.

\subsubsection{Channel Hardening}
In mMIMO systems, due to a large number of receive antennas as compared to the number of users, the Gram matrix, $\mathbf{G}=\mathbf{H}^H\mathbf{H}$, becomes diagonal dominant which is a consequence of channel hardening \cite{r1,r27}. Mathematically, it can be expressed as 
\begin{equation}
\label{eq13}
\frac {\mathbf{h}_{i}^{T}\mathbf{h}_{j}}{N_r}\rightarrow 0,\quad \forall i\neq j,~~i,j \in \left \{{1,2,\ldots ,UM}\right \},
\end{equation}
where $\mathbf{h}_j$ is the $j$th column of channel gain matrix $\mathbf{H}$. 
\subsubsection{Low-Complexity Pseudo-Inverse}
Due to diagonal dominance, we can approximate the inverse of matrix $\mathbf{G}$ by using matrix $\mathbf{D}$ \cite{r16} as 
\begin{equation}
\label{eq14}
\mathbf{G}^{-1}\approx\mathbf{D}^{-1},
\end{equation}
where $\mathbf{D}$ is the diagonal matrix containing only the diagonal entries of $\mathbf{G}$. Therefore, the problem of finding pseudo-inverse in mMIMO systems \cite{r16,r17,r18} can be approximated by using a low-complexity approximation as 
\begin{equation}
\label{eq15}
(\mathbf{H}^{H}\mathbf{H})^{-1}\mathbf{H}^{H} =  \mathbf{G}^{-1}\mathbf{H}^H \approx  \mathbf{D}^{-1}\mathbf{H}^H.
\end{equation}
It is worth noting that, the approximate inverse of matrix $\mathbf{G}$ is computed only once for each user and requires the computational complexity of $\mathcal{O}(MN_r)$.
\subsubsection{MAP Selection Rule}
Next, we use the approximate pseudo-inverse obtained in Eq. (\ref{eq15}) over the residual vector defined in Eq. \ref{eq10} to obtain the favorable MAP as
\begin{equation}
\label{eq16}
\mathbf{e}_j^{(t)} = \mathbf{W}_j \left(\mathbf{y}-\sum_{i=1, i\neq j}^{U}\mathbf{H}_i\hat{\mathbf{x}}_i^{(t-1)} \right).
\end{equation} 
where $\mathbf{W}_j=\mathbf{D}^{-1}_j\mathbf{H}_j^H$ and $\mathbf{D}_j$ is the diagonal of the Gram matrix $\mathbf{H}_j^H\mathbf{H}_j$. From $\mathbf{e}_j^{(t)}$, we select the element having the largest magnitude value, i.e., $\vert{e}_{j,i}^{(t)}\vert$ for all $i=1,2,\cdots,M$ and select $k$ as the index of the largest value which is nothing but the index of the most favorable MAP. The magnitude value of $\vert{e}_{j,i}^{(t)}\vert$ resembles the non-zero location in the transmitted symbol vector $\mathbf{x}_j$ which can be mathematically analysed using Eq. (\ref{eq16}). The vector $\mathbf{e}_j^{(t)}$ can be simplified as
\begin{eqnarray}
\nonumber
\mathbf{e}_j^{(t)} & = & \mathbf{W}_j \left(\mathbf{H}_j \mathbf{x}_j+\sum_{i=1, i\neq j}^{U} \mathbf{H}_i \mathbf{x}_i + \mathbf{n}-\sum_{i=1, i\neq j}^{U}\mathbf{H}_i\hat{\mathbf{x}}_i^{(t-1)} \right)\\
\label{eq17}
&=& \mathbf{W}_j \mathbf{H}_j \mathbf{x}_j + \sum_{i=1, i\neq j}^{U} \mathbf{W}_j \mathbf{H}_i \left(\mathbf{x}_i -\hat{\mathbf{x}}_i^{(t-1)}\right) + \mathbf{W}_j \mathbf{n}.
\end{eqnarray}
After incorporating the value of $\mathbf{W}_j$ in Eq. (\ref{eq17}) and further solving we get 
\begin{eqnarray}
\label{eq18}
&=& \mathbf{D}^{-1}_j\mathbf{G}_j \mathbf{x}_j + \sum_{i=1, i\neq j}^{U} \mathbf{D}^{-1}_j\mathbf{G}{ji} \left(\mathbf{x}_i -\hat{\mathbf{x}}_i^{(t-1)}\right) + \widetilde{\mathbf{n}},
\end{eqnarray}
where $\widetilde{\mathbf{n}}=\mathbf{D}^{-1}_j\mathbf{H}_j^H\mathbf{n}$ and $\mathbf{G}_{ji} = \mathbf{H}_j^H \mathbf{H}_i$. Due to channel hardening $\mathbf{D}^{-1}_j\mathbf{G}_j \approx \mathbf{I}$ and $\mathbf{D}^{-1}_j\mathbf{G}{ji} \approx \mathbf{O}$. Clearly, the vector $\mathbf{e}_j^{(t)} $ reduces to a combination the transmitted symbol $\mathbf{x}_j$ and the noise vector $\widetilde{\mathbf{n}}$, i.e., $\mathbf{e}_j^{(t)}=\mathbf{x}_j + \widetilde{\mathbf{n}}$. Obviously, the vector $\mathbf{x}_j$ has only one non-zero location, and therefore, the magnitude $\vert{e}_{j,i}^{(t)}\vert$ for all $i=1,2,\cdots,M$ have the maximum value corresponding to that non-zero location.\\
Finally, we decide the MBM symbol corresponding to the $j$th user as
\begin{equation}
\label{eq19}
\hat{\mathbf{s}}_j^{(t)} = [0,\cdots, 0, \underbrace{\beta}_{k\text{th coordinate}}, 0, \cdots, 0]^T 
\end{equation}
where $\beta = \mathcal{Q}[{e}_{j,k}]$ such that $\beta \in \mathbb{A}$. The symbol for the $j$th user is updated only if it results in the reduction of the norm of residual vector, i.e., $\hat{\mathbf{x}}_j^{(t)}=\hat{\mathbf{s}}_j^{(t)}$ if $\Vert\mathbf{r}^{(t)}\Vert^2 > \Vert\mathbf{r}^{(t)}+\mathbf{H}_j(\hat{\mathbf{x}}_j^{(t-1)}-\hat{\mathbf{s}}_j^{(t)})\Vert^2$. Similarly, the symbol corresponding to each user is detected in a single iteration. The algorithm is run for multiple iterations say $L$ for refining the detected symbols and obtaining the minimum norm of residual vector. Algorithm \ref{algo3} presents the pseudo code of the MAP-ISD algorithm for symbol detection in MBM-mMIMO systems.
\begin{algorithm}[H]
{
\caption{MAP-ISD Algorithm}
\label{algo3}
\small
\begin{algorithmic}[1]
\State {{\bf Input:} ${\mathbf y}$, ${\mathbf H}$, $N_r, U, n_{rf}, L $}
\State {\bf Initialization:} $\mathbf{x}^{(0)}=\mathbf{0}, \mathbf{r}^{(0)}=\mathbf{y}, \mathbf{W}_i, \forall i=1,2,\cdots,U, \ \ t = 0\ \  (\text{Iteration index})$\;
\Repeat
\State $\hat{\mathbf{x}}^{(t)}=\hat{\mathbf{x}}^{(t-1)}, \mathbf{r}^{(t)}=\mathbf{r}^{(t-1)}, \ \ t = t+1 $\;
\For {$(j=1 , ++j, j\leq U )$} 
\State Compute : $\mathbf{e}_j=\mathbf{W}_j(\mathbf{r}^{(t)}+\mathbf{H}_j\hat{\mathbf{x}}_j^{(t-1)})$\;
\State Find the element having largest  $\vert{e}_{j,i}^{(t)}\vert$ for all $i=1,2,\cdots,M $\;
\State Find the MBM symbol for the $j$th user as: $\hat{\mathbf{s}}_j^{(t)} = [0,\cdots, 0, \underbrace{\beta}_{k\text{th coordinate}}, 0, \cdots, 0]^T$ where $\beta = \mathcal{Q}[{e}_{j,k}]$ \;
\If {$\Vert\mathbf{r}^{(t)}\Vert^2 > \Vert\mathbf{r}^{(t)}+\mathbf{H}_j(\hat{\mathbf{x}}_j^{(t-1)}-\hat{\mathbf{s}}_j^{(t)})\Vert^2$}
\State Update the residual vector as $\mathbf{r}^{(t)} = \mathbf{r}^{(t)}+\mathbf{H}_j(\hat{\mathbf{x}}_j^{(t-1)}-\hat{\mathbf{s}}_j^{(t)})$\;
\State Update the symbol vector for $j$th user as $\hat{\mathbf{x}}_j^{(t)}=\hat{\mathbf{s}}_j^{(t)}$\;
\EndIf
\EndFor
\Until{$\hat{\mathbf{x}}^{(t)}=\hat{\mathbf{x}}^{(t-1)}$ or $t=L$ {(terminate the iterative processing or if no update in the solution)}}
\State {\bf Output:}$\hat{{\mathbf{x}}}_{t}$
\end{algorithmic}
}
\end{algorithm}

\subsection{$K$-favorable MAP based IIC}
In this section, we discuss the proposed KMAP-IIC algorithm for detecting symbols in MBM-mMIMO systems. Through selection of multiple favorable MAPs, the computational complexity of the existing IIC can be reduced significantly without compromising the BER performance (as discussed in Section \ref{sec4}). In the proposed algorithm, multiple favorable MAPs are selected using Eq. (\ref{eq16}) and a list $\mathbb{L}_j^{(t)}, \ \forall j=1,2,\cdots, U$, of $K$-favorable MAPs is generated. From $\mathbf{e}_j^{(t)}$, we select $K$ elements having largest magnitude value to update $\mathbb{L}_j^{(t)}$ i.e. we sort $\vert{e}_{j,i}^{(t)}\vert$ for all $i=1,2,\cdots,M$ in descending order, and select the indices of first $K$ element. \\
Next, we define the reduced size set $\tilde{\mathbb{S}}_{MBM}^{j,(t)}$ of favorable MBM symbol vectors for the $j$th user in the $t$th iteration as
\begin{gather}
\nonumber
\tilde{\mathbb{S}}_{MBM}^{j,(t)}=\{\mathbf{s}_{k,i} \in \mathbb{A}_0 : k=\mathbb{L}_j^{(t)}(1),\mathbb{L}_j^{(t)}(2),\cdots,\mathbb{L}_j^{(t)}(K),\\
\nonumber  i=1,2,\cdots,\vert\mathbb{A}\vert\}\\
\label{eq20} 
\text{s.t.}\ \ \mathbf{s}_{k,i}=[0,\cdots, 0, \underbrace{\alpha_i}_{k\text{th coordinate}}, 0, \cdots, 0]^T, \alpha_i\in\mathbb{A}.
\end{gather} 
After finding the list of favorable MAPs and the reduced size set of MBM symbol vectors, a search is performed over the set $\tilde{\mathbb{S}}_{MBM}^{j,(t)}$ for all the users in order to determine the best candidate solution corresponding to each user. The solution in the set  $\tilde{\mathbb{S}}_{MBM}^{j,(t)}$ which minimizes the norm of residual vector is selected as the best candidate solution for the $j$th user defined as
\begin{equation}
\label{eq21}
\hat{\mathbf{s}}_j^{(t)} = \arg\min_{\tilde{\mathbf{s}}_q\in \tilde{\mathbb{S}}_{MBM}^{j,(t)}} \Vert{\mathbf{r}_j^{(t)}-\mathbf{H}_j\tilde{\mathbf{s}}_q}\Vert^2.
\end{equation}
Moreover, for a square- and rectangular-QAM constellation, a low-complexity search, proposed in \cite{r22}, is utilized in the proposed algorithm to reduce the computational complexity further. The real and imaginary parts of the estimated solution for square- and rectangular-QAM constellations can be expressed mathematically as
\begin{gather}
\nonumber
\Re(\hat{\mathbf{s}}_l)=\min \left[ \max \left(2 \lfloor \frac{m_1 + 1}{2} \rceil -1, - N_1 +1\right), N_1 - 1 \right],\\
\label{eq22}
\Im(\hat{\mathbf{s}}_l)=\min \left[ \max \left(2 \lfloor \frac{m_2 + 1}{2} \rceil -1, - N_2 +1\right), N_2 - 1 \right],
\end{gather} 
where $m_1=\Re({z}_l)$, $m_2=\Im({z}_l)$ and $z_l=\frac{\mathbf{h}_{j,l}^H \mathbf{r}_j^{(t)}}{\Vert\mathbf{h}_{j,l}\Vert^2}$, and $N_1,N_2$ are the sizes two PAMs on the real and imaginary axis of the QAM constellation, respectively. Thus the best possible solution corresponding to the $l$th MAP of the $j$th user is $\hat{\mathbf{s}}_l=\Re(\hat{\mathbf{s}}_l)+i\Im(\hat{\mathbf{s}}_l)$ \cite{r22}. This makes the search independent of the number of points in the constellation set, and therefore, reduces the computational complexity incurred during the detection. \\
The search for the best possible solution is followed by symbol update stage where symbol vectors corresponding to the selected user is updated. A greedy update strategy proposed in \cite{r8} is used which requires $U$ iterations to update the solution corresponding to several users in order to minimize the ML cost. In each iteration, the user for which the ML cost is selected by using
\begin{equation}
\label{eq23}
v=\arg\min_{i=1,2,\cdots,U}\Vert\mathbf{r}^{(t)}+\mathbf{H}_i(\hat{\mathbf{x}}_i^{(t)}-\hat{\mathbf{s}}_i^{(t)})\Vert^2,
\end{equation}
and the solution corresponding to the selected user is updated. This update strategy is initiated only after completion of search over reduced size set for all the users. Algorithm \ref{algo4} presents the pseudo code of the KMAP-IIC algorithm.

\begin{algorithm}[H]
{
\caption{KMAP-IIC Algorithm}
\label{algo4}
\small
\begin{algorithmic}[1]
\State {{\bf Input:} ${\mathbf y}$, ${\mathbf H}$, $N_r, U, n_{rf}, L, K $}
\State {\bf Initialization:} $\mathbf{x}^{(0)}=\mathbf{0}, \mathbf{r}^{(0)}=\mathbf{y}, \mathbf{W}_i, \forall i=1,2,\cdots,U, \ \ t = 0\ \  (\text{Iteration index})$\;
\Repeat
\State $\hat{\mathbf{x}}^{(t)}=\hat{\mathbf{x}}^{(t-1)}, \mathbf{r}^{(t)}=\mathbf{r}^{(t-1)}, \ \ t =t+1$\;
\For {$(j=1 , ++j, j\leq U )$} 
\State Compute : $\mathbf{e}_j=\mathbf{W}_j(\mathbf{r}^{(t)}+\mathbf{H}_j\hat{\mathbf{x}}_j^{(t-1)})$\;
\State Sort the elements in Descending Order and update $\mathbb{L}_j^{(t)}$ i.e. select indices of first $K$ elements from $\arg sort( \vert{e}_{j,1}^{(t)}\vert, \vert{e}_{j,2}^{(t)}\vert,\hdots, \vert{e}_{j,M}^{(t)}\vert, descend)$\;
\State Generate the Reduced Search Space $\tilde{\mathbb{S}}_{MBM}^{j,(t)}$\;
\State Perform search for the $j$th user as: $\hat{\mathbf{s}}_j^{(t)} = \arg\min_{\tilde{\mathbf{s}}_q\in \tilde{\mathbb{S}}_{MBM}^{j,(t)}} \Vert{\mathbf{r}_j^{(t)}-\mathbf{H}_j\tilde{\mathbf{s}}_q}\Vert^2$\;
\EndFor
\For {$(j=1 , ++j, j\leq U )$}
\State Select the first element of $v=\arg\min_{i=1,2,\cdots,U}\Vert\mathbf{r}^{(t)}+\mathbf{H}_i(\hat{\mathbf{x}}_i^{(t)}-\hat{\mathbf{s}}_i^{(t)})\Vert^2$\;
\If {$\Vert\mathbf{r}^{(t)}\Vert^2 > \Vert\mathbf{r}^{(t)}+\mathbf{H}_v(\hat{\mathbf{x}}_v^{(t)}-\hat{\mathbf{s}}_v^{(t)})\Vert^2$}
\State Update the residual vector as $\mathbf{r}^{(t)} = \mathbf{r}^{(t)}+\mathbf{H}_v(\hat{\mathbf{x}}_v^{(t)}-\hat{\mathbf{s}}_v^{(t)})$\;
\State Update the symbol vector for $v$th user as $\hat{\mathbf{x}}_v^{(t)}=\hat{\mathbf{s}}_v^{(t)}$\;
\Else
\State break\;
\EndIf 
\If {$\hat{\mathbf{x}}^{(t)}=\hat{\mathbf{x}}^{(t-1)}$}
\State break\; 
\EndIf
\EndFor
\Until{$\hat{\mathbf{x}}^{(t)}=\hat{\mathbf{x}}^{(t-1)}$ or $t=L$ {(terminate the iterative processing or if no update in the solution)}}
\State {\bf Output:}$\hat{{\mathbf{x}}}_{t}$
\end{algorithmic}
}
\end{algorithm}

\section{Simulation results}
\label{sec4}
In this section, we present the simulation results of the proposed detection algorithms in terms of BER performance and the computational complexity for MBM-mMIMO systems. We also compare the BER performance of the proposed algorithms with the performance of MMSE detector \cite{r7}, IESP algorithm \cite{r7} and IIC algorithm \cite{r8}. We consider $128\times 16$ and $128\times 20$ MBM-mMIMO systems with $n_{rf}=3$, $n_{rf}=4$, $n_{rf}=5$ and $n_{rf}=6$, respectively. In summary, Table \ref{tab1a} shows list of the parameters used in simulation of BER performance and computational complexity.
\begin{table}[H]
\caption{List of parameters used in simulation.}
\label{tab1a}   
\centering    
\begin{tabular}{lll}
\hline\noalign{\smallskip}
 Figure &  MBM-mMIMO System & Parameters Used \\
\noalign{\smallskip}\hline\noalign{\smallskip}
Fig. 2 & $N_r = 128$, $U=20$, $n_{rf}=3$ & 4-QAM, $L=1,2,4,6,8$ \\
Fig. 3 &$N_r = 128$, $U=16$, $n_{rf}=4$ & 4-QAM, $L=1,2,4,6,8$ \\
Fig. 4 & $N_r = 128$, $U=20$, $n_{rf}=3$ & 4-QAM, $L=1,2,4,6$, $K=M/2$ \\
Fig. 5 & $N_r = 128$, $U=20$, $n_{rf}=3$ & 4-QAM, $L=6$, $K=1, M/4, M/2$ \\
Fig. 6 & $N_r = 128$, $U=20$, $n_{rf}=4$ & 4-QAM, $L=6$, $K=1, M/4, M/2$ \\
Fig. 7 & $N_r = 128$, $U=16$, $n_{rf}=6$ & 4-QAM, $L=6$, $K=1, M/4, M/2$ \\
Fig. 8 & $N_r = 128$, $U=20$, $n_{rf}=3$ & 16-QAM, $L=6$, $K=1, M/4, M/2$ \\
Fig. 9 & $N_r = 128$, $U=20$, $n_{rf}=5$ & 16-QAM, $L=6$, $K=1, M/4, M/2$ \\
Fig. 10 & $U=20$, $n_{rf}=4$ & 4-QAM, $L=6$, $K=1, M/4, M/2$ \\
Fig. 11 & $N_r=128$, $U=16,20$, $n_{rf}=3,4$ & 4-QAM, $L=6$, $K=1, M/4, M/2$, SNR = 5 dB \\
Fig. 12 & $U=16$, $n_{rf}=4$ & 4-QAM, $L=6$, $K=1, M/4, M/2$, SNR = 5 dB \\
\noalign{\smallskip}\hline
\end{tabular}
\end{table}

\subsection{Bit Error Rate Performance}
\label{ssec41}
\begin{figure}[H]
	\centering 
	\includegraphics[width=3.5in]{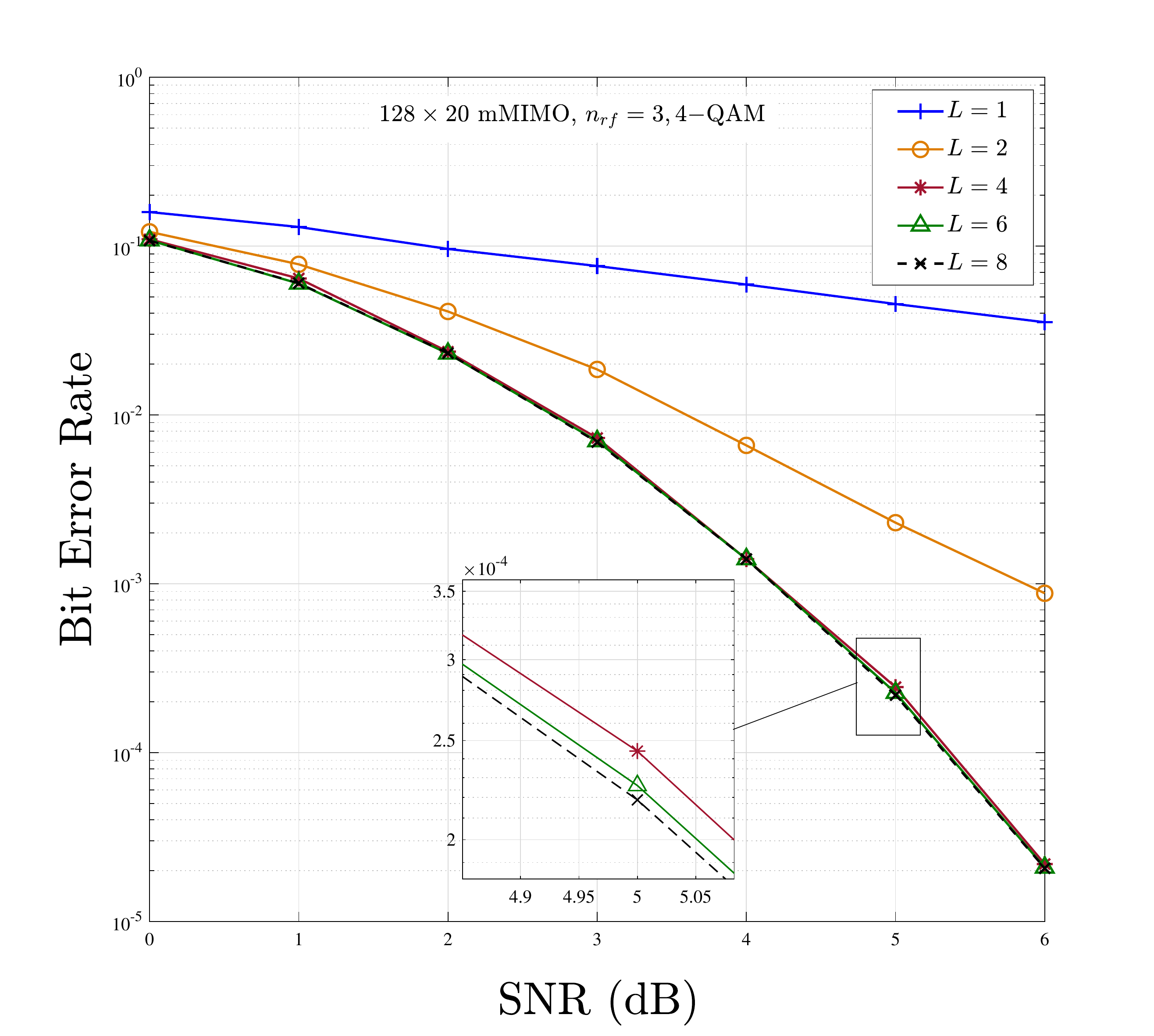}
	\caption{BER performance of MAP-ISD for $128\times 20$, $n_{rf}=3$ MBM-mMIMO system with 4-QAM modulation.}
    \label{fig1}
\end{figure}

\begin{figure}[H]
	\centering 
	\includegraphics[width=3.5in]{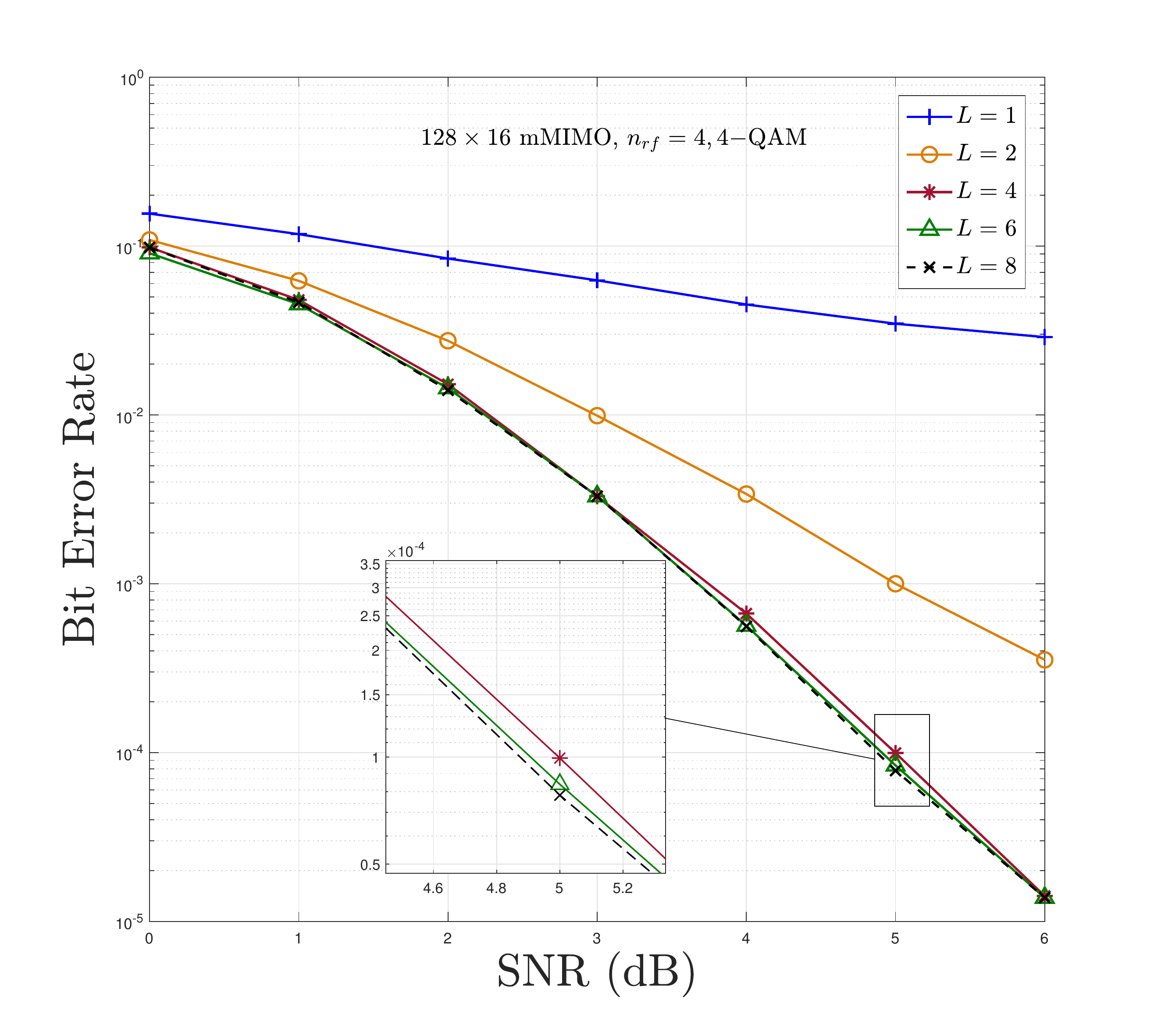}
	\caption{BER performance of MAP-ISD for $128\times 16$, $n_{rf}=4$ MBM-mMIMO system with 4-QAM modulation.}
    \label{fig2}
\end{figure}

In Figs. \ref{fig1} and \ref{fig2}, we present the BER performance of MAP-ISD with different iterations $L=1,2,4,6$ and $8$ for $128\times 20$ mMIMO system, $n_{rf} =3$ and $128\times 16$ mMIMO system, $n_{rf}=4$, respectively. It is observed that the BER performance of MAP-ISD improves with increase in the number of iterations and converges after $L=6$. Significant improvement in BER performance is observed for increase in the value of $L$ from 1 to 2 and from 2 to 4 whereas marginal improvement is observed for further increase in the value of $L$.  
\begin{figure}[H]
	\centering 
	\includegraphics[width=3.5in]{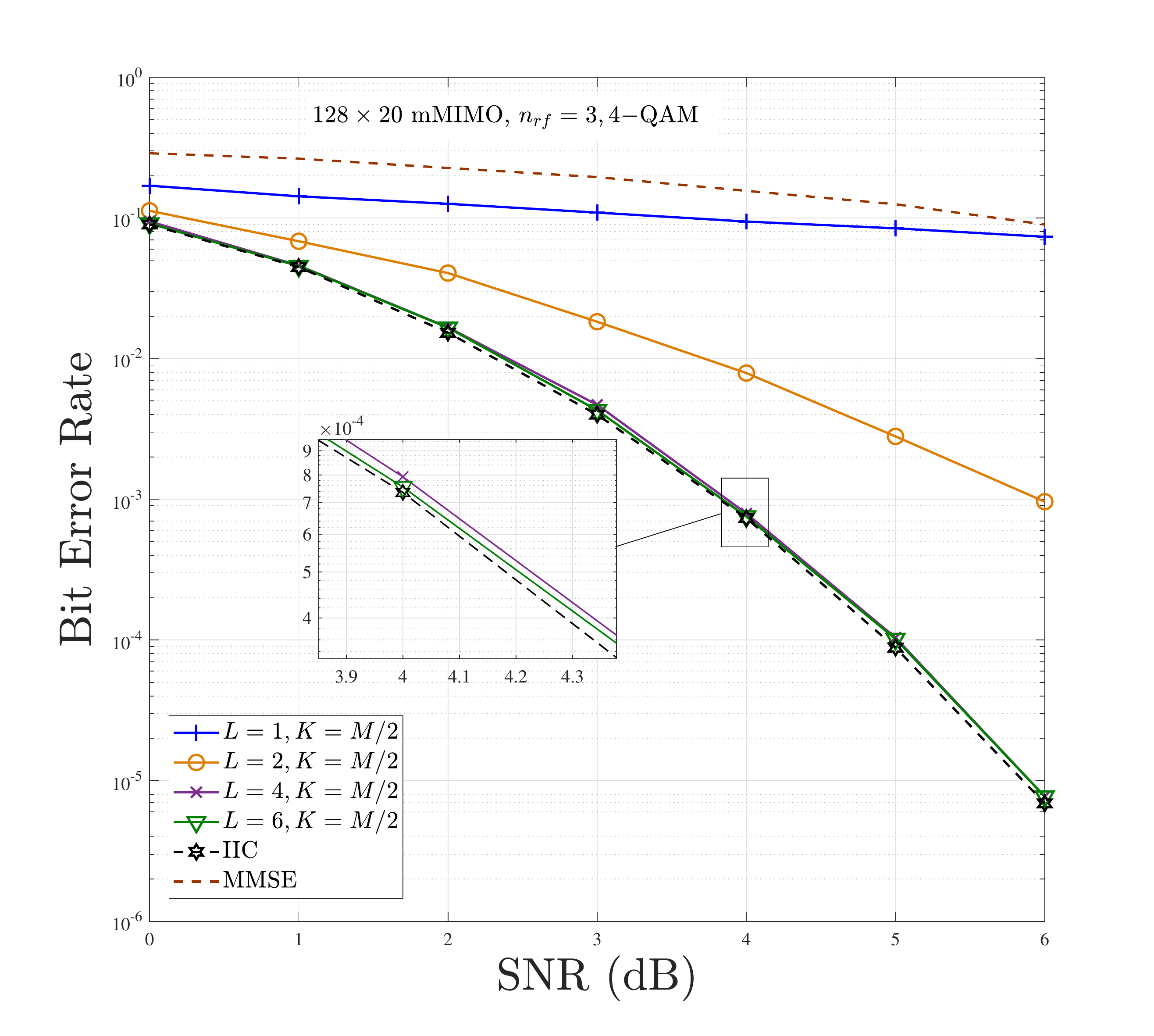}
	\caption{BER performance of KMAP-IIC with $K=M/2$ for $128\times 20$, $n_{rf}=4$ MBM-mMIMO system with 4-QAM modulation.}
    \label{fig3}
\end{figure}
In Fig. \ref{fig3}, we simulate the BER performance of KMAP-IIC algorithm with different iterations $L =1,2,4$ and $6$ for $128\times 20$ mMIMO systems with 4-QAM, $K=M/2$ and $n_{rf} =3$ MBM. Similar observations can be drawn which suggests that the BER performance of the KMAP-IIC algorithm converges for $L=6$. Therefore, for comparison of BER of the proposed algorithms, i.e. MAP-ISD and KMAP-IIC, with other detection techniques (discussed later in the section) we use $L=6$ iterations.
\begin{figure}[H]
	\centering 
	\includegraphics[width=3.5in]{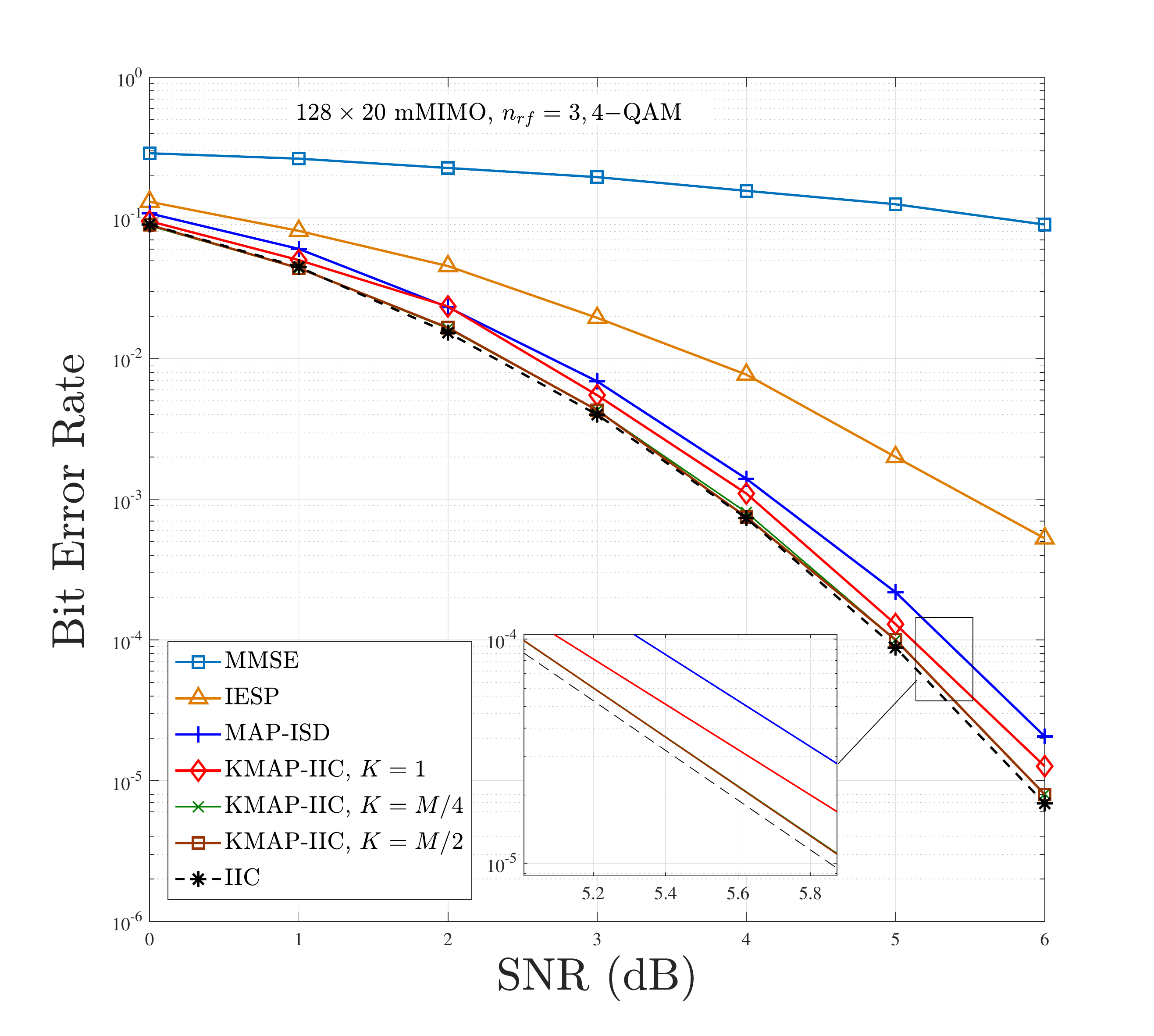}
	\caption{BER performance comparison for $128\times 20$, $n_{rf}=3$ MBM-mMIMO system with 4-QAM modulation.}
    \label{fig4}
\end{figure}
\indent In Fig. \ref{fig4}, we compare the BER performance of the proposed algorithms with other algorithms such as IIC, IESP and MMSE for $128\times 20$ mMIMO with 4-QAM, $n_{rf}=3$ MBM system. For comparison of KMAP-IIC, we consider three different values for $K$, i.e. $K=1$, $K=M/4$ and $K=M/2$. It is observed that, MAP-ISD and KMAP-IIC algorithms outperform the MMSE and the IESP algorithms. Moreover, the KMAP-IIC achieves BER performance close to within $0.1$ dB to that of the IIC algorithm with $K=M/2$. For a target BER of $10^{-4}$, MAP-ISD achieves performance within 0.5 dB of the IIC algorithm. Therefore, due to significant low-computational complexity of MAP-ISD (discussed later in Section \ref{ssec42}), it would be a better choice over other detection techniques with marginal loss in performance. Fig. \ref{fig5} shows the comparison of BER for $128\times 20$ mMIMO with 4-QAM, $n_{rf}=4$ MBM systems. It is observed that in contrast to Fig. \ref{fig4}, the BER performance gap between MAP-ISD and KMAP-IIC increases with increase in $n_{rf}$ from 3 to 4. The key reason behind such performance degradation is the error propagation in interference cancellation of MAP-ISD which increases with increase in $n_{rf}$. It is interesting to note that, KMAP-IIC with $K=1$ achieves BER performance close to within 0.1 dB to that of IIC, KMAP-IIC with $K=M/4$ and KMAP-IIC with $K=M/2$. Therefore, in MBM-mMIMO systems with moderate values of $n_{rf}$, KMAP-IIC with $K=1$ could be a better choice. \\
 \begin{figure}[H]
	\centering 
	\includegraphics[width=3.5in]{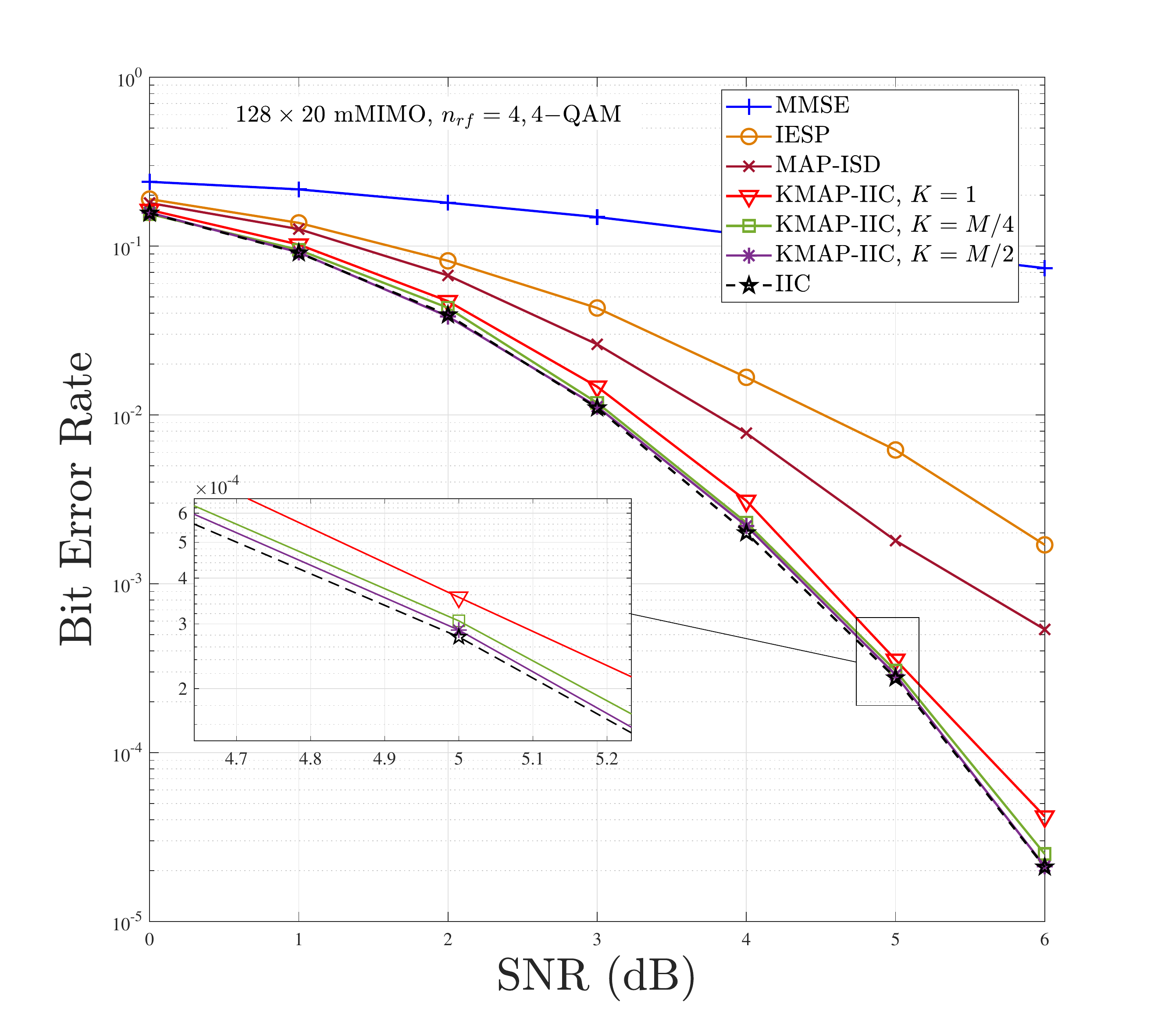}
	\caption{BER performance comparison for $128\times 20$, $n_{rf}=4$ MBM-mMIMO system with 4-QAM modulation.}
    \label{fig5}
\end{figure}
\indent Fig. \ref{fig6} presents the BER performance comparison for $128\times 20$ mMIMO with 4-QAM, $n_{rf}=6$ MBM system. It is observed that the performance of KMAP-IIC with $K=M/2$ is same as that of the IIC algorithm. The performance of KMAP-IIC with $K=M/4$ and $K=1$ is close to within $0.2$ dB and $0.5$ dB, respectively, of the performance of IIC algorithm. However, the performance of MAP-ISD is far inferior as compared to the KMAP-IIC, IIC and IESP algorithms. Clearly, from Figs. \ref{fig4}, \ref{fig5} and \ref{fig6}, the performance of MAP-ISD degrades as compared to the IIC algorithm which makes MAP-ISD less selective for systems having high values of $n_{rf}$. However, in such scenarios KMAP-IIC still performs up to mark. Therefore, from the aforementioned BER performance results, it can be concluded that the choice of low-complexity (discussed in next subsection) reasonable detection technique between MAP-ISD and KMAP-IIC depends significantly on the value of $n_{rf}$ RF mirrors and the number of uplink users.
\begin{figure}[H]
	\centering 
	\includegraphics[width=3.5in]{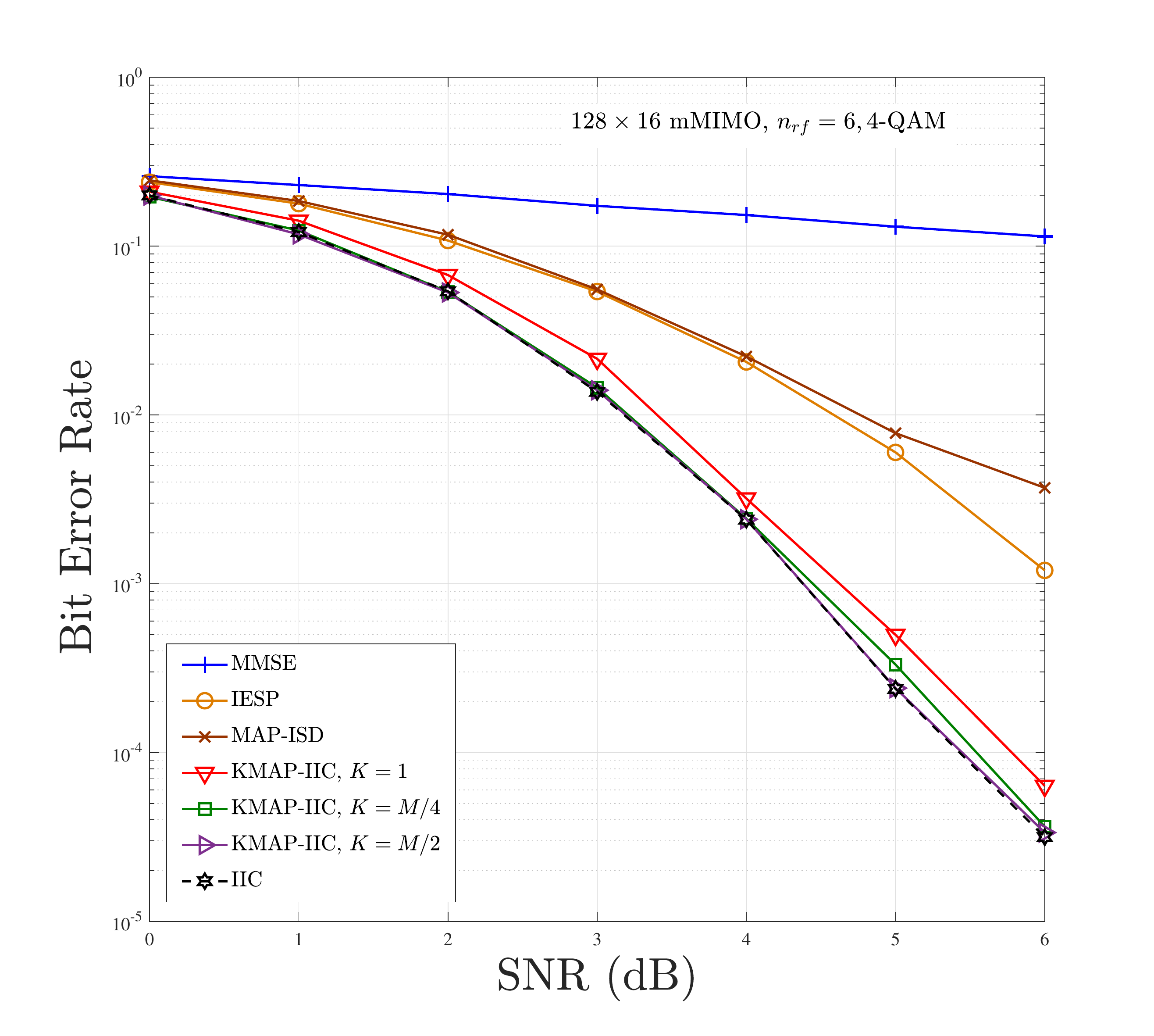}
	\caption{BER performance comparison for $128\times 16$, $n_{rf}=6$ MBM-mMIMO system with 4-QAM modulation.}
    \label{fig6}
\end{figure}
\begin{figure}[H]
	\centering 
	\includegraphics[width=3.5in]{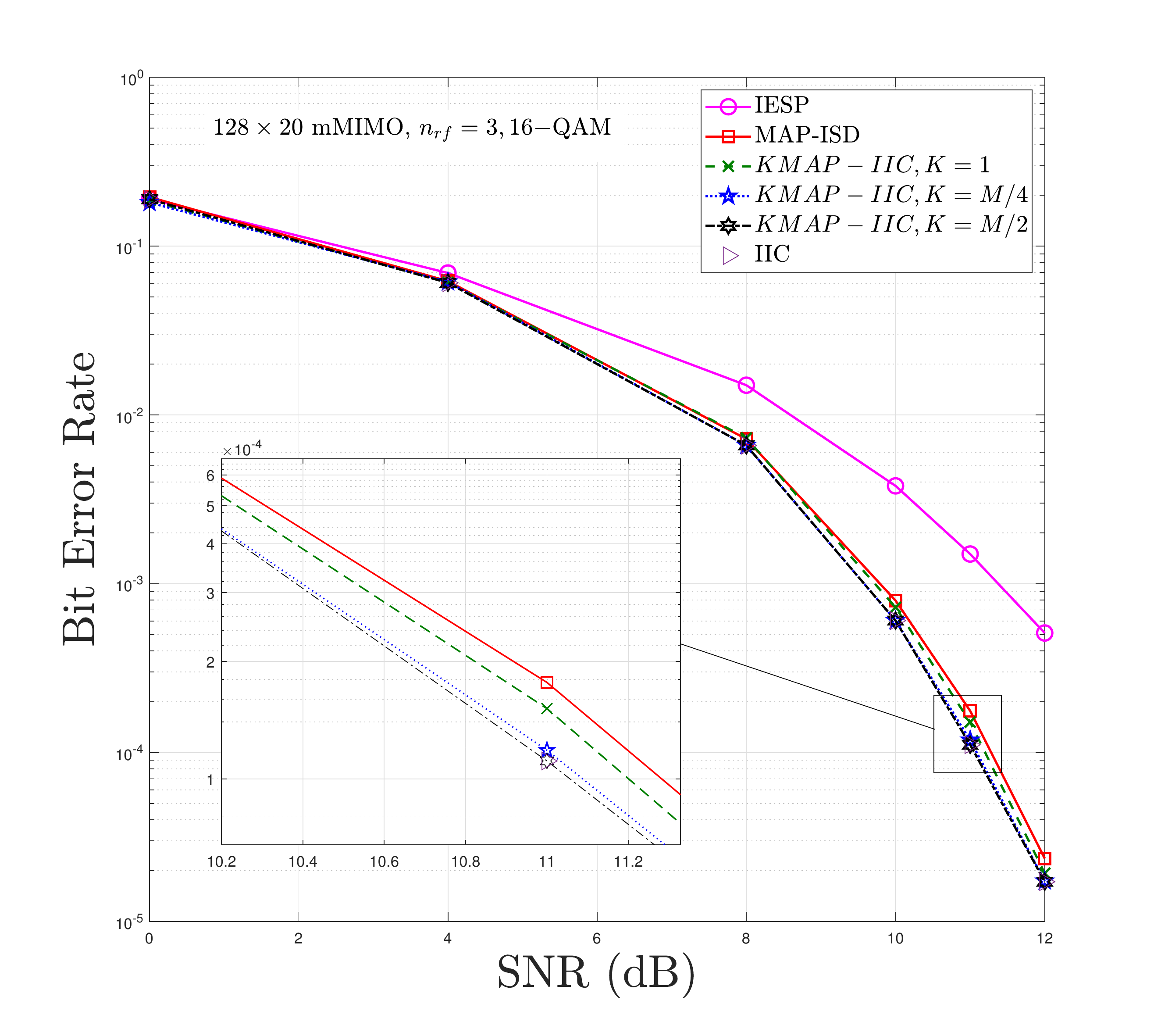}
	\caption{BER performance comparison for $128\times 20$, $n_{rf}=3$ MBM-mMIMO system with 16-QAM modulation.}
    \label{fig7}
\end{figure}

\begin{figure}[H]
	\centering 
	\includegraphics[width=3.5in]{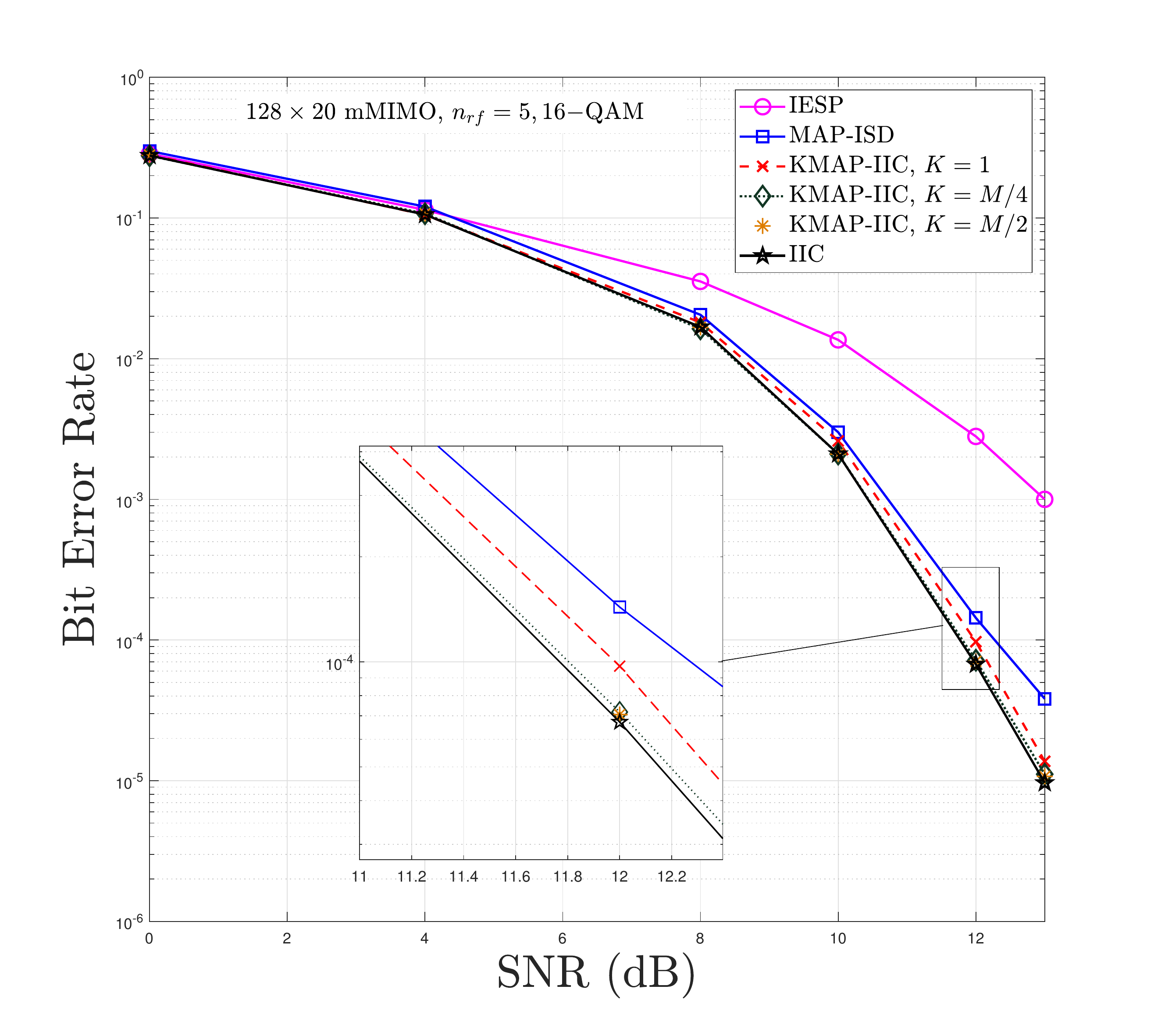}
	\caption{BER performance comparison for $128\times 20$, $n_{rf}=5$ MBM-mMIMO system with 16-QAM modulation.}
    \label{fig8}
\end{figure}
\begin{figure}[H]
	\centering 
	\includegraphics[width=3.5in]{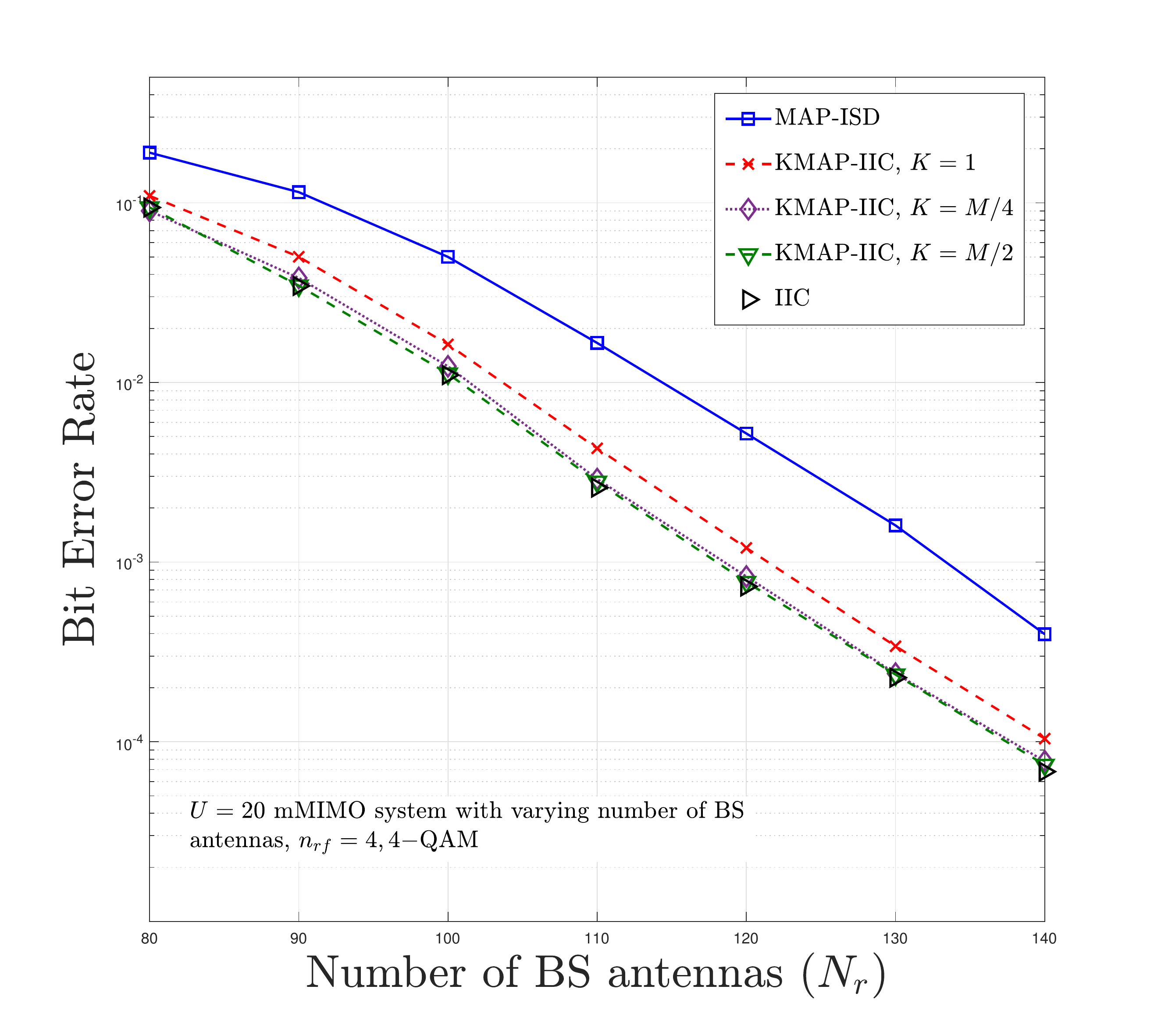}
	\caption{BER performance comparison for $U=20$ mMIMO with $n_{rf}=4$ and 4-QAM modulation for different values of $N_r$.}
    \label{fig9}
\end{figure}

 In Figs. \ref{fig7} and \ref{fig8}, the BER performance comparison is performed for $128 \times 20$ mMIMO system with $n_{rf} =3$ and $n_{rf} = 5$ with 16-QAM modulation. Observations reveal that the proposed algorithms perform superior over the IESP algorithm. However, the performance of MAP-ISD and KMAP-IIC with $K=1$ degrades with an increase in $n_{rf}$ as compared to KMAP-IIC with $K=M/4$, KMAP-IIC with $K=M/2$ and IIC. It is also observed that KMAP-IIC with $K=M/4$ achieves performance close to KMAP-IIC with $K=M/2$ and IIC algorithms.\\
 \indent In Fig. \ref{fig9}, we present the BER performance with variation in the number of BS antennas from $N_r=80$ to $N_r=140$ for $U=20$, $n_{rf}=4$ and 4-QAM MBM-mMIMO system. It is observed that the BER performance improves with an increase in the BS antennas. Also, the performance of KMAP-IIC with different values of $K$, i.e., $K=1, M/4, M/2$, is close to that of the IIC algorithm. On the other hand, the performance of MAP-ISD is far from IIC. For example, for a target BER of $5 \times 10^{-4}$, required BS antennas for IIC and KMAP-IIC are around $125$, whereas MAP-ISD requires around $140$ BS antennas. Moreover, channel-hardening is not observed to degrade the performance compared to the IIC algorithm.

\subsection{Computational Complexity}
\label{ssec42}
\begin{figure}[H]
	\centering 
	\includegraphics[width=3.5in]{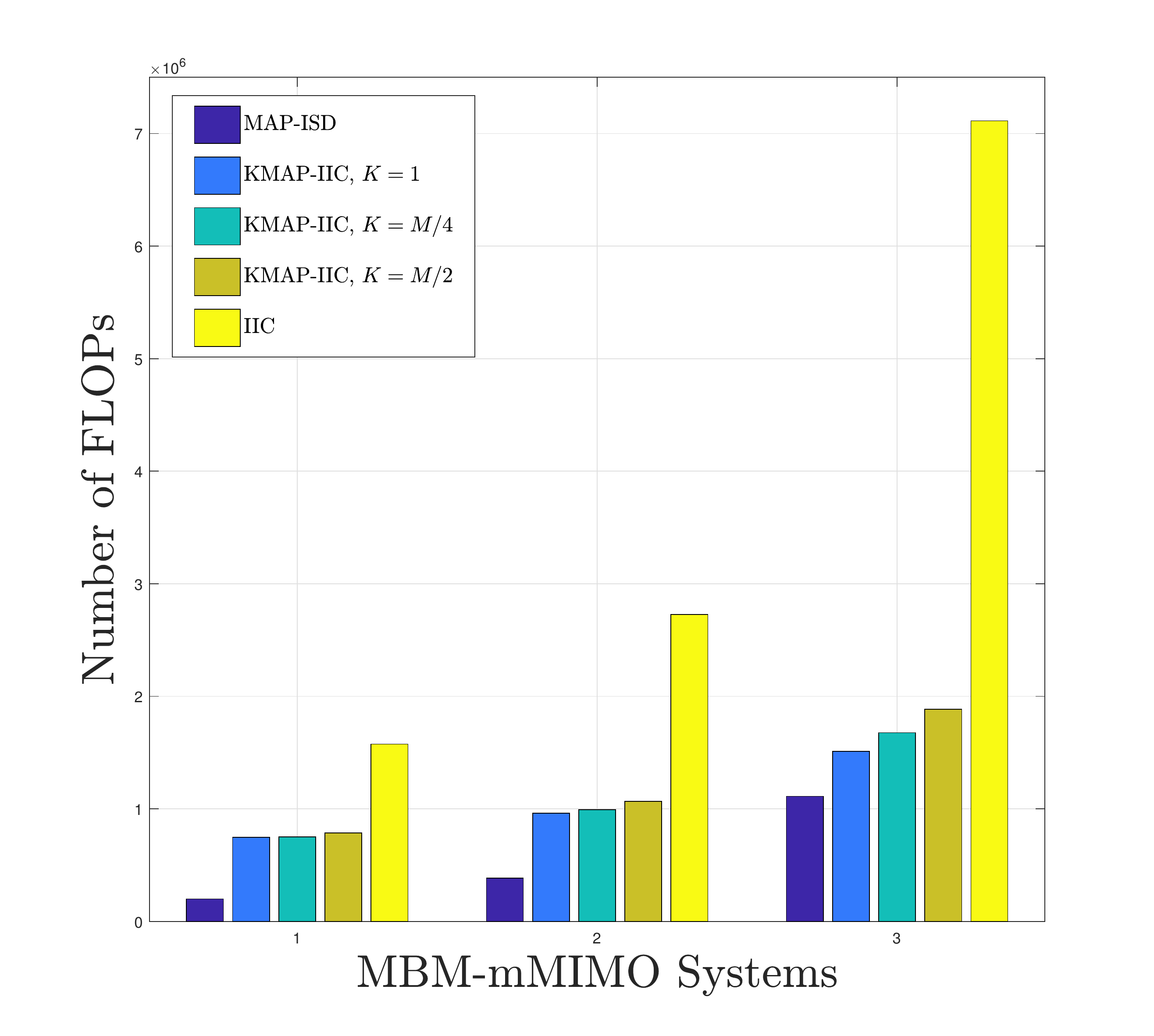}
	\caption{Computational complexity comparison of different detection techniques for different MBM-mMIMO systems.}
    \label{fig10}
\end{figure}

This section discusses the computational complexity of the proposed algorithms and compares them with the IIC algorithm \cite{r8}. First, we discuss the approximate computations involved in different steps of the IIC algorithm in Table \ref{tab1} in terms of floating-point operations (FLOPs) \cite{r23},\cite{r24}. It is to note that the algorithms' initialization does not require any computation due to all zero initial solutions.
\begin{table}[H] 
\caption{Approximate number of computations in each iteration of IIC algorithm}
\label{tab1}  
\centering   
\begin{tabular}{lll}
\hline\noalign{\smallskip}
 Steps involved in IIC &  Computational complexity \\
\noalign{\smallskip}\hline\noalign{\smallskip}
Interference cancellation & $2N_r $ \\
Performing ML search & $(4N_r-1)MU \vert\mathbb{A}\vert $ \\
Finding the solution using greedy search & $(6N_r-1)U^2 $  \\
\noalign{\smallskip}\hline
\end{tabular}
\end{table}
In Table \ref{tab2} and \ref{tab3}, we present the computations required in different steps of the MAP-ISD and KMAP-IIC algorithms, respectively.
\begin{table}[H]
\caption{Approximate number of computations in each iteration of MAP-ISD}
\label{tab2}   
\centering    
\begin{tabular}{ll}
\hline\noalign{\smallskip}
 Steps involved in MAP-ISD &  Computational complexity \\
\noalign{\smallskip}\hline\noalign{\smallskip}
Finding low-complexity matrix inverse (required only once) & $(2N_r-1)MU$ \\ 
Finding the most favorable MAP & $2MN_rU$ \\
Finding reliability of the solution & $(4N_r-1)U$\\
\noalign{\smallskip}\hline
\end{tabular}
\end{table}
\begin{table}[H]
\caption{Approximate number of computations in each iteration of KMAP-IIC}
\label{tab3}   
\centering    
\begin{tabular}{ll}
\hline\noalign{\smallskip}
 Steps involved in KMAP-IIC &  Computational complexity \\
\noalign{\smallskip}\hline\noalign{\smallskip}
Finding low-complexity matrix inverse (required only once) & $(2N_r-1)MU$ \\ 
Finding favorable MAPs & $2MN_rU$ \\
Performing low-complexity search & $(2N_r+4)KU$ \\
Finding the solution using greedy search & $(6N_r-1)U^2 $  \\
\noalign{\smallskip}\hline
\end{tabular}
\end{table}
It is observed from Tables \ref{tab1}, \ref{tab2}, and \ref{tab3} that the computations in MAP-ISD are significantly less than that of IIC and KMAP-IIC algorithm, which is due to the complex search mechanism in IIC and KMAP-IIC algorithm. The computation of finding a low-complexity matrix inversion in MAP-ISD and KMAP-IIC is required only once. Therefore, the reduction in search space due to favorable MAP selection and low-complexity search are the key factors for the overall reduction in the proposed algorithms' computational complexity. Fig. \ref{fig10} presents the comparison of average number of FLOPs for different MBM-mMIMO systems at SNR= 5 dB. In Fig. \ref{fig10}, System 1 refers to $128 \times 20$ mMIMO systems with $n_{rf}=3$, system 2 refers to $128 \times 20$ mMIMO systems with $n_{rf}=4$ and system 3 refers to $128 \times 16$ mMIMO systems with $n_{rf}=6$. Observations reveal that KMAP-IIC with $K=M/2$ achieves significant computational gain of upto $70 \%$ and $63 \%$ for $128 \times 16$, $n_{rf}=6$ and $128 \times 20$, $n_{rf}=4$ systems, respectively. This gain increases further to $76 \%$ and $80 \%$ when MAP-ISD algorithm is used for $128 \times 16$, $n_{rf}=6$ and $128 \times 20$, $n_{rf}=4$ systems, respectively. It is clear from the figure that the computational complexity of MAP-ISD is significantly less as compared to the KMAP-IIC with $K=1$, $K=M/4$ and $K=M/2$. Therefore, in case of MBM-mMIMO systems with less value of $n_{rf}$, MAP-ISD is suitable over KMAP-IIC with $K=1$ and other algorithms. On the other hand for moderate values of $n={rf}$ KMAP-IIC with $K=1$ and $K=M/4$ can be used to achieve a reasonable BER performance.   
\begin{figure}[H]
	\centering 
	\includegraphics[width=3.5in]{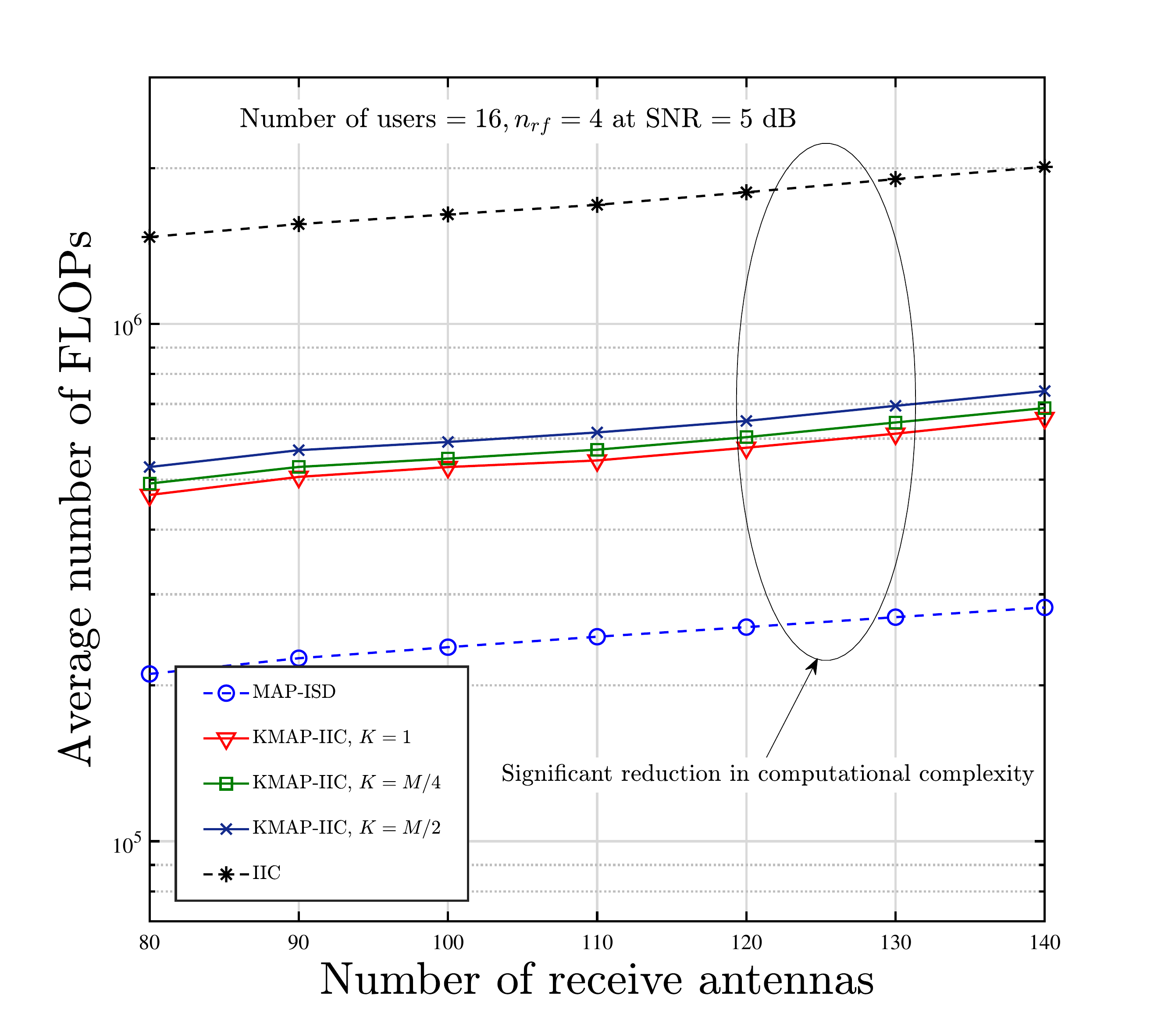}
	\caption{Computational complexity comparison of different detection techniques versus the number of receive antennas.}
    \label{fig11}
\end{figure}
In Fig. \ref{fig11}, we present the comparison on the average number of floating point operations (FLOPs) of the IIC algorithm with MAP-ISD and KMAP-IIC algorithms, respectively, with respect to the number of receive antennas for MBM-mMIMO system having $16$ users and $n_{rf}=4$ at each user. Observation reveals that the computational complexity increase linearly with increase in the number of receive antennas at the station. It is also observed that the computational complexity increases with increase in $K$. Furthermore, the complexity of MAP-ISD is significantly less compared with the KMAP-IIC algorithm and the IIC algorithm \cite{r8}.

\section{Conclusion}
\label{sec5} 
We proposed low complexity interference cancellation based algorithms for symbol detection in MBM-mMIMO systems. First, we proposed a MAP selection based ISD algorithm which detects symbols in a sequential manner while nullifying interference from all other users. Then, we presented an approach to reduce the search space in the IIC algorithm and devised the KMAP-IIC algorithm. The key idea was to introduce a metric and a selection rule for selecting the list of the favorable MAPs in MBM corresponding to each user. The proposed algorithms achieve superior performance-complexity trade-off over the existing detection techniques in MBM-mMIMO systems. Observations reveal that when the number of users and the RF mirrors are comparatively less MAP-ISD algorithm achieves similar performance with almost $75\%$ and $87\%$ savings in computational complexity as compared to KMAP-IIC and IIC algorithm, respectively. However, with an increase in the number of users and the RF mirrors, KMAP-IIC delivers better BER performance over MAP-ISD with almost $65\%$ to $70 \%$ savings in computational complexity for different value of $K$ over the IIC algorithm. 

\begin{acknowledgements}
This work is supported by the Start-up Research Grant (file no. SRG/2019/000654) scheme of Science and Engineering Research Board, Department of Science and Technology, Government of India.
\end{acknowledgements}

\section*{Conflict of interest}
The authors declare that they have no conflict of interest.

\end{document}